\documentstyle[12pt]{article}
\input epsf
\def\pct#1{ \centerline{ \epsfbox{#1.eps}}}

\def\8{\infty}
\def\oh{\frac{1}{2}}
\def\ot{\frac{1}{3}}
\def\oq{\frac{1}{4}}

\def\tq{\frac{3}{4}}
\def\d{\partial}
\def\i{\imath\,}
\def\alp{\alpha}
\def\bet{\beta}
\def\gam{\gamma}
\def\lam{\lambda}
\def\eps{\epsilon}

\def\del{\delta}
\def\dal{\partial_{\alpha}}

\def\undertext#1{\vtop{\hbox{#1}\kern 1pt \hrule}}
\def\ra{\rightarrow}
\def\lra{\longrightarrow}

\def\VEV#1{\left\langle #1\right\rangle}
\def\tr{\hbox{tr}\,}

\def\pp#1{\frac{\partial}{\partial#1}}
\def\pbyp#1#2{\frac{\partial#1}{\partial#2}}

\def\fbyf#1#2{\frac{\delta#1}{\delta#2}}

\def\br{\\ \nonumber &&}
\def\inv#1{\frac{1}{#1}}
\def\be{\begin{equation}}
\def\ee{\end{equation}}
\def\bea{\begin{eqnarray} &&}
\def\eea{\end{eqnarray}}
\def\ct#1{\cite{#1}}
\def\rf#1{(\ref{#1})}
\def\EXP#1{\exp\left(#1\right)}

\def\1N{$\frac{1}{N}$ expansion}

\def\et{{\cal E}}

\def\C{\oint_C d}
\def\Y{\Psi_C(\gamma)}

\def\NS{Navier-Stokes }
\def\val{v_{\alpha}}
\def\vbe{v_{\beta}}

\def\ral{r_{\alpha}}
\def\rbe{r_{\beta}}
\def\rga{r_{\gamma}}

\begin{document}

\begin{titlepage}

{\bf October '93}\hfill	  {\bf PUPT-1426}\\

\begin{center}

{\bf LOOP EQUATION AND AREA LAW IN TURBULENCE}

\vspace{1.5cm}

{\bf   A.A.~Migdal}

\vspace{1.0cm}

{\it  Physics Department, Princeton University,\\
Jadwin Hall, Princeton, NJ 08544-1000.\\
E-mail: migdal@acm.princeton.edu}

\vspace{1.9cm}

\end{center}

\abstract{
This is the extended version of the preprint \ct{Loop}, based on the lectures
given in Cargese Summer School and Chernogolovka Summer School in 93. The
incompressible fluid dynamics is reformulated as dynamics of closed loops $C$
in coordinate space. We derive explicit functional equation for the pdf of the
circulation $P_C(\Gamma)$ which allows the scaling solutions  in inertial range
of spatial scales. The pdf decays as exponential of some power  of $
\Gamma^3/A^2 $  where $A$ is the minimal area inside the loop.
}
\vfill
\end{titlepage}

\tableofcontents

\section{Introduction}

Incompressible fluid dynamics underlies the vast majority of natural phenomena.
It is described by famous Navier-Stokes equation
\begin{equation}
\dot{v}_{\alpha} = \nu \partial_{\beta}^2 v_{\alpha} - v_{\beta}
\partial_{\beta} v_{\alpha} - \partial_{\alpha} p \\;\;
\partial_{\alpha}v_{\alpha} = 0 \label{eq1}
\end{equation}
which is nonlinear, and therefore hard to solve.
This nonlinearity makes life more interesting, though, as it leads to
turbulence. Solving this equation with appropriate initial and boundary
conditions we expect to obtain the chaotic behavior of velocity field.

The simplest boundary conditions correspond to infinite space with
vanishing velocity at infinity. We are looking for the translation
invariant probability distribution for velocity field, with infinite range of
the wavelengths. In order to compensate for the energy dissipation, we add the
usual random force to the \NS equations, with the short wavelength support,
corresponding to large scale energy pumping.

One may attempt to describe this probability distribution by the
Hopf generating functional (the angular bracket denote time  averaging, or
ensemble averaging over realizations of the random forces)
\begin{equation}
Z[J] = \left \langle \exp \left( \int d^3 r
J_{\alpha}(r)v_{\alpha}(r)\right)
\right \rangle \label{eq2}
\end{equation}
which is known to satisfy linear functional differential equation
\begin{equation}
\dot{Z} = H\left[J,\frac{\delta}{\delta J} \right] Z
\label{eq3}
\end{equation}
similar to the Schr\"odinger equation for Quantum Field Theory, and
equally hard to solve. Nobody managed to go beyond the Taylor
expansion in source $ J $ , which corresponds to the obvious chain of equations
for the equal time correlation functions of velocity
field in various points in space. The same equations could be obtained
directly from Navier-Stokes equations, so the Hopf equation looks
useless.

In this work\footnote{see also \ct{Loop} where this approach was initiated and
\ct{cells} where its relation with the generalized Hamiltonian dynamics and the
Gibbs-Boltzmann statistics was established} we argue, that one could
significantly simplify the Hopf functional without loosing information about
correlation functions.
This simplified functional depends upon the set of 3 periodic
functions of one variable
\begin{equation}
C : r_{\alpha} = C_{\alpha}(\theta)\\;\; 0< \theta< 2\pi
\end{equation}
which set describes the closed loop in coordinate space. The
correlation functions reduce to certain functional derivatives of our loop
functional with respect to $ C(\theta)$ at vanishing loop $ C \rightarrow 0 $.

The properties of the loop functional at large loop $ C $ also have
physical significance. Like the Wilson loops in Gauge Theory, they
describe the statistics of large scale structures of vorticity field, which is
analogous to the gauge field strength.

In Appendix A we recover the expansion in inverse powers of viscosity by direct
iterations of the loop equation.

In Appendix  B we study the matrix formulation of the \NS equation, which may
serve as a basis of the random matrix description of turbulence.

In Appendix  C we study the reduced dynamics, corresponding to the
functional Fourier transform of the loop functional. We argue, that
instead of 3D \NS equations one can use the 1D equations for the Fourier loop
$P_{\alp}(\theta,t)$.

In Appendix D we discuss the relation between the initial data for
velocity field and the $P$ field, and we find particular realisation for these
initial data in terms of the gaussian random variables.

In Appendix E we introduce the generating functional for the scalar
products $ P_{\alp}(\theta)P_{\alp}(\theta') $. The advantage of this
functional over the original $\Psi[C]$ functional is the smoother continuum
limit.

In Appendix F we discuss the possible numerical implementations
of the reduced loop dynamics.

In Appendix G we show uniqueness of the tensor area law within certain class of
functionals.

In Appendix H we present the modern view at the old problem of the minimal
surface.

In Appendix I we show that the triple Kolmogorov correlation function
corresponds to a vanishing correlation of vorticity with two velocity fields.

\section{The Loop Calculus}

We suggest to use in turbulence the following version of the Hopf functional
\begin{equation}
\Psi \left[C \right] = \left \langle \exp
\left(
    \frac{\i }{\nu}  \oint dC_{\alpha}(\theta)
v_{\alpha}\left(C(\theta)\right) \right) \right \rangle \label{eq4}
\end{equation}
which we call the loop functional or the loop field.
It is implied that all angular variable $\theta$ run from $ 0 $ to $ 2\pi$and
that all the functions of this variable are $ 2\pi$
periodic.\footnote{This parametrization of the loop is a matter of
convention, as the loop functional is parametric invariant.} The viscosity $
\nu $ was inserted in denominator in exponential, as the only parameter of
proper dimension. As we shall see below, it plays the role, similar to the
Planck's constant in Quantum mechanics, the turbulencecorresponding to the WKB
limit $ \nu \rightarrow 0 $. \footnote{One could also insert any numerical
parameter in exponential, but this factor could be eliminated by space- and/or
time rescaling.}

As for the imaginary unit $\i$, there are two reasons to insert it in the
exponential. First, it makes the motion compact: the phase factor goes  around
the unit circle, when the velocity field fluctuates. So, at large  times one
may expect the ergodicity, with well defined average functional  bounded by $1$
by absolute value. Second, with this factor of $\i$, the  irreversibility of
the problem is manifest. The time reversal corresponds  to the complex
conjugation of $\Psi$, so that imaginary part of the  asymptotic value of
$\Psi$ at $t \ra \8$ measures the effects of  dissipation.

The loop orientation reversal $ C(\theta) \ra C(2\pi - \theta) $ also  leads to
the complex conjugation,  so it is equivalent to the time  reversal.  This
symmetry implies, that any  correlator of odd/even number  of velocities should
be integrated odd/even number of times over the loop,  and it must enter with
an imaginary/real factor. Later, we shall use this  property in the area law.

We shall often use the field theory notations for the loop integrals,
\be
\Psi \left[C \right] = \left \langle \exp
\left(
    \frac{\i }{\nu}  \oint_C dr_{\alpha}v_{\alpha}
 \right) \right \rangle \label{eq4'}
\ee
This loop integral can be reduced to the surface
integral of vorticity field
\be
\omega_{\mu\nu} = \d_{\mu}v_{\nu}-\d_{\nu}v_{\mu}
\ee
by the Stokes theorem
\begin{equation}
\Gamma_C[v] \equiv \oint_C dr_{\alpha}v_{\alpha}= \int_{S}  d
\sigma_{\mu\nu} \omega_{\mu\nu} \\;\; \partial S = C
\end{equation}

This is the well-known velocity circulation, which measures the net
strength of the vortex lines, passing through the loop $ C $.  Would we fix
initial loop $ C $ and let it move with the flow, the loop field would be
conserved by the Euler equation, so that only the viscosity effects would be
responsible for its time evolution. However, this is not what we are trying to
do.  We take the Euler rather than Lagrange dynamics, so that the loop is fixed
 in space, and hence $\Psi$ is time dependent already in the Euler  equations.
The difference between Euler and \NS equations is the time  irreversibility,
which leads to complex average $\Psi$ in \NS dynamics.

It is implied that this field $\Psi\left[C\right]$ is invariant under
translations of the loop $ C(\theta) \rightarrow C(\theta)+ const $. The
asymptotic behavior at large time  with proper random forcing reaches  certain
fixed point, governed by the translation- and scale invariant equations,  which
we derive in this paper.

The general Hopf functional (\ref{eq2}) reduces for the loop field for the
following  imaginary singular source
\begin{equation}
J_{\alpha}(r) =   \frac{\i }{ \nu} \oint_C dr'_{\alpha} \delta^3
\left(r'-r \right) \label{eq8}
\end{equation}

The $\Psi$  functional involves connected correlation functions of the  powers
of circulation at equal times.
\be
\Psi[C] = \EXP{\sum_{n=2}^{\8}\frac{\i^{n}}{n!\,\nu^{n}}\,\VEV{\VEV{
\Gamma_C^n[v]}}}
\ee
This expansion goes in powers of effective Reynolds number, so it diverges  in
turbulent region. There, the opposite WKB approximation will be used.

Let us come back to the general case of the arbitrary Reynolds number.  What
could be the use of such restricted Hopf functional? At first glance it seems
that we lost most of information, described by the Hopf functional, as the
general Hopf source $J$ depends upon 3 variables $ x,y,z $ whereas the loop $C$
depends of only one parameter $ \theta $. Still, this information can be
recovered by taking the loops of the singular shape, such as two infinitesimal
loops $R_1, R_2 $,  connected by a couple of wires

\pct{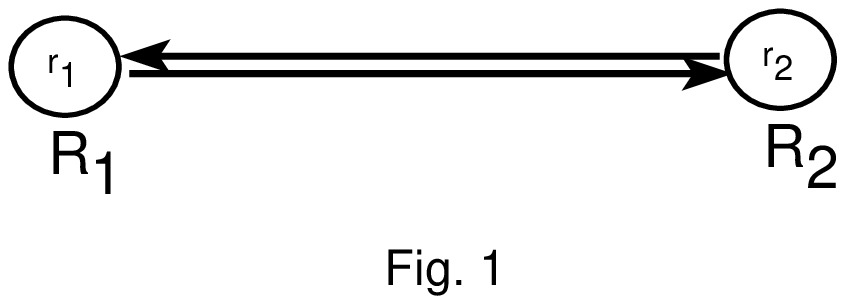}

The loop field in this case reduces to
\begin{equation}
\Psi \left[C \right] \rightarrow \left \langle  \exp
\left(
    \frac{\i}{ 2\nu} \Sigma_{\mu\nu}^{R_1}\omega_{\mu\nu}(r_1)
+\frac{\i}{ 2\nu}   \Sigma_{\mu\nu}^{R_2 } \omega_{\mu\nu}(r_2)
\right) \right \rangle
\end{equation}
where
\begin{equation}
\Sigma_{\mu\nu}^R = \oint_R d r_{\nu}r_{\mu}
\end{equation}
is the tensor area inside the loop $R$. Taking functional derivatives with
respect to the shape of $R_1$ and $R_2$ prior to shrinking them to  points, we
can bring down the product of vorticities at  $r_1$ and $r_2$.  Namely, the
variations yield
\be
\delta\Sigma_{\mu\nu}^R=  \oint _R\left(d r_{\nu}\delta r_{\mu}+ r_{\mu}d
\delta r_{\nu} \right) =  \oint_R \left( d r_{\nu}\delta r_{\mu} -d
r_{\mu}\delta r_{\nu} \right)
\ee
where integration by parts was used in the second term.

One may introduce the area derivative $\fbyf{}{\sigma_{\mu\nu}(r)}$, which
brings down the vorticity at the given point $ r $ at the loop.
\begin{equation}
-\nu^2 \frac{\delta^2 \Psi \left[C \right]}
{\delta \sigma_{\mu\nu}(r_1)\delta \sigma_{\lambda \rho}(r_2)}
\ra \left \langle \omega_{\mu\nu}(r_1)
\omega_{\lambda \rho}(r_2) \right \rangle
\end{equation}

The careful definition of these area derivatives are or paramount  importance
to us. The corresponding loop calculus was developed in\cite{Mig83} in the
context of the gauge theory. Here we rephrase
and further refine the definitions and relations established in that
work.

The basic element of the loop calculus is what we suggest to call the spike
derivative, namely the operator which adds the infinitesimal $
\Lambda $ shaped spike to the loop
\begin{equation}
D_{\alpha}(\theta,\epsilon) = \int_{\theta}^{\theta+2\epsilon}d \phi
\left(
1-\frac{\left|\theta +\epsilon - \phi\right|}{\epsilon }
\right)
    \frac{\delta}{\delta C_{\alpha}(\phi)}
\end{equation}
The finite spike operator
\begin{equation}
\Lambda(r,\theta,\epsilon) =
\exp \left( r_{\alpha}  D_{\alpha}(\theta,\epsilon) \right)
\end{equation}
adds the spike of the height $r$. This is the straight line from $
C(\theta) $ to $ C(\theta + \epsilon) + r$, followed by another
straight line from $ C(\theta+\epsilon)+r $ to $ C(\theta+2
\epsilon)$,

\pct{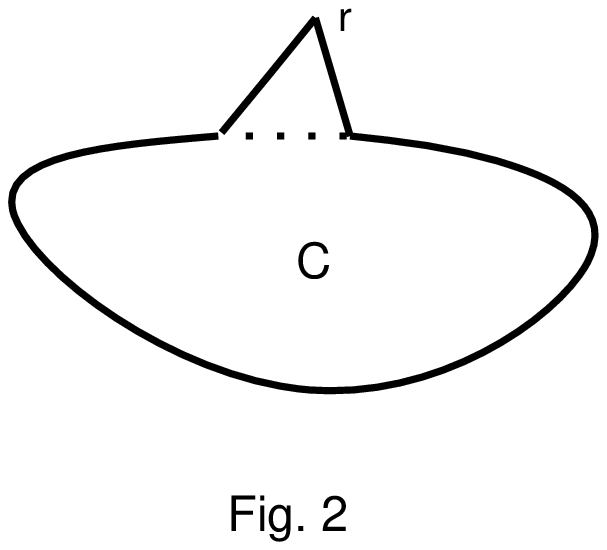}
Note, that the loop remains
closed, and the slopes remain finite, only the second derivatives
diverge. The continuity and closure of the loop eliminates the
potential part of velocity; as we shall see below,
this is necessary to obtain the loop equation.

In the limit $ \epsilon \rightarrow 0 $ these spikes are invisible, at
least for the smooth vorticity field, as one can see from the Stokes
theorem (the area inside the spike goes to zero as $ \epsilon $).
However, taking certain derivatives prior to the limit $ \epsilon
\rightarrow 0 $ we can obtain the finite contribution.

Let us consider the operator
\begin{equation}
\Pi \left(r,r',\theta ,\epsilon \right) =
\Lambda  \left(r, \theta,\frac{1}{2} \epsilon \right) \Lambda
\left(r',\theta,\epsilon \right)
\end{equation}
By construction it inserts the smaller spike on top of a bigger one,
in such a way, that a polygon appears

\pct{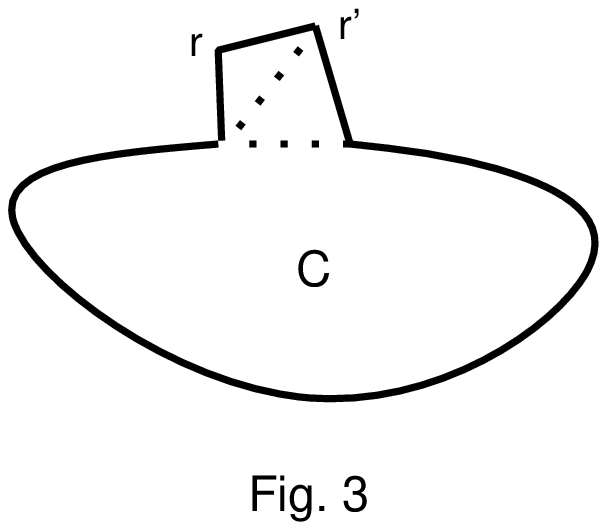}
Taking the derivatives with respect to the  vertices of
this polygon $ r, r' $ , setting $r=r'=0$ and
antisymmetrising, we find the tensor operator
\begin{equation}
\Omega_{\alpha\beta}(\theta,\epsilon) =
-\i \nu  D_{\alpha}\left(\theta,\frac{1}{2} \epsilon \right)
D_{\beta}\left(\theta,\epsilon \right) - \{\alpha \leftrightarrow\beta\}
\label{OM}
\end{equation}
which brings down the vorticity, when applied to the loop field
\begin{equation}
\Omega_{\alpha\beta}(\theta,\epsilon) \Psi \left[C \right]
\stackrel{\epsilon \rightarrow 0}{\longrightarrow}
\omega_{\alpha\beta}\left(C(\theta)\right)\Psi \left[C \right] \label{eqom}
\end{equation}

The quick  way to check these formulas is to use formal functional
derivatives
\begin{equation}
\frac{\delta \Psi \left[C \right]}{\delta C_{\alpha}(\theta)} =
C'_{\beta}(\theta) \fbyf{\Psi \left[C
\right]}{\sigma_{\alp\bet}\left(C(\theta)\right)}
\end{equation}
Taking one more functional derivative derivative we find the term with
vorticity times first derivative of the $ \delta $ function, coming from
the variation of $ C'(\theta) $
\be
\frac{\delta^2 \Psi [C ]}{\delta C_{\alp}(\theta) \delta
C_{\bet}(\theta')} = \del'(\theta-\theta')\fbyf{\Psi \left[C
\right]}{\sigma_{\alp\bet}\left(C(\theta)\right)} +
C'_{\gam}(\theta) C'_{\lam}(\theta')\frac{\delta^2\Psi \left[C
\right]}{\delta \sigma_{\alp\gam}\left(C(\theta)\right) \delta
\sigma_{\bet\lam}\left(C(\theta')\right)}
\ee
This term is the only one, which survives the limit $ \epsilon
\rightarrow 0 $ in our relation (\ref{eqom}).

So, the area derivative can be defined from the antisymmetric tensor part
of  the second functional derivative as the coefficient in front of $
\delta'(\theta-\theta') $ .  Still, it has all the properties of the first
functional derivative, as it can also be defined from the above first
variation.
The advantage of dealing with spikes is the control over the limit $\eps
\ra 0$ , which might be quite singular in applications.

So far we managed to insert the vorticity at the loop $ C $ by
variations of the loop field. Later we shall need the vorticity off
the loop, in arbitrary point in space. This can be achieved by the
following combination of the spike operators
\begin{equation}
\Lambda \left(r,\theta,\epsilon \right) \Pi
\left(r_1,r_2,\theta+\epsilon,\delta \right) \\;\; \delta \ll \epsilon
\end{equation}
This operator inserts the $ \Pi $ shaped little loop at the top of the
bigger spike, in other words, this little loop is translated by a
distance $r$ by the big spike.

Taking derivatives, we find the operator of finite translation of the
vorticity
\begin{equation}
\Lambda \left(r,\theta,\epsilon \right)
\Omega_{\alpha\beta}(\theta+ \epsilon ,\delta)
\end{equation}
and the corresponding infinitesimal translation operator
\begin{equation}
D_{\mu}(\theta,\epsilon)\Omega_{\alpha\beta}(\theta+ \epsilon ,\delta)
\end{equation}
which inserts $ \partial_{\mu} \omega_{\alpha \beta} \left( C(\theta)
\right) $ when applied to the loop field.

Coming back to the correlation function, we are going now to construct
the operator, which would insert two vorticities separated by a distance.
Let us note that the global $ \Lambda $ spike
\begin{equation}
\Lambda \left(r,0,\pi \right) = \exp
\left(
    r_{\alpha}\int_{0}^{2\pi}d
\phi  \left(1- \frac{ \left|\phi-\pi \right|}{\pi} \right)
\frac{\delta}{\delta C_{\alpha}(\phi)}\right)
\end{equation}
when applied to a  shrunk loop $ C(\phi) = 0 $ does nothing but
the backtracking from $0$ to $r$

\pct{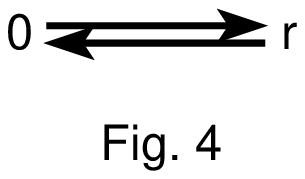}
This means that the operator
\begin{equation}
\Omega_{\alpha\beta}(0 ,\delta)\Omega_{\lambda \rho}(\pi ,\delta)
\Lambda \left(r,0,\pi \right)
\end{equation}
when applied to the loop field for a shrunk loop yields the vorticity
correlation function
\begin{equation}
\Omega_{\alpha\beta}(0 ,\delta)\Omega_{\lambda \rho}(\pi ,\delta)
\Lambda \left(r,0,\pi \right) \Psi [0] = \left \langle \omega_{\alpha
\beta}(0) \omega_{\lambda \rho}(r) \right \rangle
\end{equation}

The higher correlation functions of vorticities could be constructed in a
similar fashion, using the spike operators. As for the velocity, one
should solve the Poisson equation
\begin{equation}
\partial_{\mu}^2 v_{\alpha}(r) = \partial_{\beta} \omega_{\beta \alpha}(r)
\end{equation}
with the proper boundary conditions , say, $ v=0 $ at infinity.
Formally,
\begin{equation}
v_{\alpha}(r) =
\frac{1}{\partial_{\mu}^{2}}\partial_{\beta} \omega_{\beta \alpha}(r)
\end{equation}

This suggests the following formal definition of  the velocity
operator
\begin{equation}
V_{\alpha}(\theta,\epsilon,\delta) = \frac{1}{D_{\mu}^2(\theta,\epsilon)}
D_{\beta}(\theta,\epsilon) \Omega_{\beta \alpha}(\theta,\delta)\\;\;
\delta \ll \epsilon
\label{VOM}
\end{equation}
\begin{equation}
V_{\alpha}(\theta,\epsilon,\delta)\Psi[C] \stackrel{\delta,\epsilon
\rightarrow 0}{\longrightarrow} v_{\alpha} \left(C(\theta) \right) \Psi[C]
\end{equation}

Another version of this formula is the following integral
\begin{equation}
V_{\alpha}(\theta,\epsilon,\delta)= \int d^3  \rho
\frac{\rho_{\beta}}{4 \pi |\rho|^3}\Lambda \left(\rho,\theta,\epsilon
\right)
\Omega_{\alpha\beta}(\theta+ \epsilon ,\delta)
\end{equation}
where the $ \Lambda $ operator shifts the  $ \Omega $ by a distance $
\rho $ off the original loop at the point $ r = C(\theta + \epsilon)
$

\pct{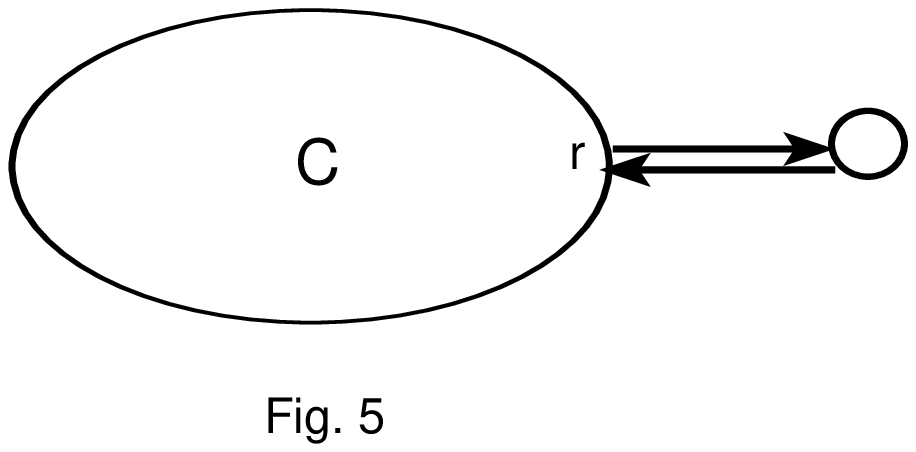}

\section{Loop Equation}

Let us now derive exact equation for the loop functional.
Taking the time derivative of the original definition, and using the
Navier-Stokes equation we get in front of exponential
\begin{equation}
\oint_C d r_{\alpha}  \frac{\i}{ \nu}
\left(
    \nu \partial_{\beta}^2 v_{\alpha} - v_{\beta}
\partial_{\beta} v_{\alpha} - \partial_{\alpha} p \right)
\end{equation}
The term with the pressure gradient yields zero after integration over
the closed loop, and the velocity gradients in the first two terms
could be expressed in terms of vorticity up to irrelevant gradient
terms, so that we find
\begin{equation}
\oint_C d r_{\alpha}  \frac{\i}{ \nu}
\left(
    \nu \partial_{\beta} \omega_{\beta \alpha} - v_{\beta}
\omega_{\beta \alpha}
\right) \label{Orig}
\end{equation}

Replacing the vorticity and velocity  by the operators discussed in the
previous Section we find the following loop equation (in explicit
notations)
\bea
-\i\dot{\Psi}[C] = \br
\oint d C_{\alpha}(\theta)
\left(
    D_{\beta}(\theta,\epsilon) \Omega_{\beta \alpha}(\theta,\epsilon) +
\frac{1}{ \nu}
\int d^3  \rho
\frac{\rho_{\gam}}{4 \pi |\rho|^3}\Lambda \left(\rho,\theta,\epsilon
\right)
\Omega_{\gam\bet}(\theta+ \epsilon ,\delta)\Omega_{\beta
\alpha}(\theta,\delta) \right) \Psi[C]
\label{PsiC}
\eea

The more compact form of this equation, using the notations of
\cite{Mig83}, reads
\bea
\i\,\nu\dot{\Psi}[C] = {\cal H}_C\Psi \br
{\cal H}_C \equiv \nu^2\oint_{C} dr_{\alpha}
\left(
    \i\partial_{\beta} \frac{\delta }{\delta \sigma_{\beta \alpha}(r)}+
\int d^3 r'\frac{r'_{\gamma}-r_{\gam}}{4 \pi |r-r'|^3}
\frac{\delta^2}{\delta \sigma_{\beta \alpha}(r)
\delta \sigma_{\beta \gamma}(r')}
\right)
\label{OLD}
\eea
Now we observe that viscosity $ \nu $ appears in front of time  and
spatial derivatives, like the Planck constant $\hbar$ in Quantum
mechanics. Our loop hamiltonian ${\cal H}_C$ is not hermitean, due to
dissipation. It contains the second loop derivatives, so it represents a
(nonlocal!) kinetic term in loop space.

So far, we considered so called decaying turbulence, without external
energy source. The  energy
\be
E = \int d^3 r \oh \, \val^2
\ee
would eventually all dissipate, so that the fluid would stop. In this case
the loop wave function $\Psi$ would asymptotically approach $1$
\be
\Psi[C] \stackrel{t \ra \8}\lra 1
\ee

In order to reach the steady state, we  add to the right side of the \NS
equation the usual gaussian random forces $f_{\alp}(r,t)$ with the space
dependent correlation function
\be
\VEV{f_{\alp}(r,t)f_{\bet}(r',t')} = \delta_{\alp\bet}\delta(t-t')F(r-r')
\ee
concentrated at at small wavelengths, i.e. slowly varying with $r-r'$.

Using the identity
\be
\VEV{f_{\alp}(r,t) \Phi[v(.)]} = \int d^3 r' F(r-r')
\fbyf{\Phi[v(.)]}{\val(r')}
\ee
which is valid for arbitrary functional $\Phi$  we find the following
imaginary potential term in the loop hamiltonian
\be
\delta{\cal H}_C \equiv \i\,U[C]= \frac{\i}{\nu}\,\oint_{C}
dr_{\alpha}\oint_{C} dr'_{\alpha} F(r-r')
\ee

Note, that  orientation reversal together with complex conjugation changes
the sign of the loop hamiltonian, as it should. The potential part
involves two loop integrations times imaginary constant. The first term in
the kinetic part has one loop integration, one loop derivative times
imaginary constant. The second kinetic term has one loop integration, two
loop derivatives and real constant. The left side of the loop equation has
no loop integrations, no loop derivatives, but has a factor of $\i$.

The relation between the potential and kinetic parts of the loop
hamiltonian depends of viscosity, or, better to say, it depends upon the
Reynolds number, which is the ratio of the typical circulation to
viscosity. In the viscous limit, when the Reynolds number is small, the
loop wave function is close to $1$. The perturbation expansion in $
\inv{\nu}$ goes in powers of the potential, in the same way, as in Quantum
mechanics. The second (nonlocal) term in kinetic part of the hamiltonian
also serves as a small perturbation (it corresponds to nonlinear term in
the \NS equation).
The first term of this perturbation expansion is just
\be
\Psi[C] \ra 1 - \int \frac{d^3
k}{(2\pi)^3}\frac{\tilde{F}(k)}{2\nu^3\,k^2} \left|\oint_C d \ral e^{\i k
r}\right|^2
\ee
with $\tilde{F}(k)$ being the Fourier transform of $F(r)$.
This term is real, as it corresponds to the two-velocity correlation. The
next term comes from the triple correlation of velocity, and this term is
purely imaginary, so that the dissipation shows up.

This expansion can be derived  by direct iterations in the loop space as
in \cite{Mig83}, inverting the operator in the local part of the kinetic
term in the hamiltonian.  This expansion is discussed in Appendix A. The
results agree with the straightforward iterations of  the \NS equations in
powers of the random force, starting from zero velocity.

So, we have the familiar situation, like in QCD, where the perturbation
theory breaks because of the infrared divergencies. For arbitrarily small
force, in a large system, the region of small $k$ would yield large
contribution to the terms of the perturbation expansion. Therefore, one
should take the opposite WKB limit $\nu \ra 0$.

In this limit, the wave function should behave as the usual WKB wave
function, i.e. as an exponential
\be
\Psi[C] \ra \EXP{\frac{\i\,S[C]}{\nu}}
\ee
The effective loop Action $S[C]$ satisfies the loop space Hamilton-Jacobi
equation
\be
\dot{S}[C] =-\i U[C] + \oint_{C} dr_{\alpha}
\int d^3 r'\frac{r'_{\gamma}-r_{\gam}}{4 \pi |r-r'|^3}
\frac{\delta S}{\delta \sigma_{\beta \alpha}(r)}
\frac{\delta S}{\delta \sigma_{\beta \gamma}(r')}
\label{SC}
\ee
The imaginary part of $S[C]$ comes from imaginary potential $U[C]$, which
distinguishes our theory from the reversible Quantum mechanics. The sign
of $\Im S$ must be positive definite,  since $ |\Psi| <1$. As for  the
real part of $S[C]$, it changes the sign under the loop orientation
reversal $C(\theta) \ra C(2\pi-\theta) $.

At finite viscosity there would be an additional term
\be
-\nu\oint_{C} dr_{\alpha}\partial_{\beta} \frac{\delta S[C]}{\delta
\sigma_{\beta\alpha}(r)}
-\i\nu \oint_{C} dr_{\alpha}
\int d^3 r' \frac{r'_{\gamma}-r_{\gam}}{4 \pi |r-r'|^3}
\frac{\delta^2 S[C]}{\delta \sigma_{\beta \alpha}(r) \delta \sigma_{\beta
\gamma}(r')}
\ee
on the right of \rf{SC}.  As for the term
\be
-\oint_{C} dr_{\alpha}
    \i\left(\partial_{\beta}S[C]\right) \frac{\delta S[C]}{\delta
\sigma_{\beta \alpha}(r)}
\ee
which formally arises in the loop equation, this term vanishes, since
$\partial_{\beta}S[C]=0$. This operator inserts backtracking at some point
at the loop without first applying the loop derivative at this point. As
it was discussed in the previous Section, such backtracking  does not
change the loop functional. This issue was discussed at length in
\ct{Mig83}, where the Leibnitz rule for the operator $ \dal
\fbyf{}{\sigma_{\bet\gam}} $ was established
\be
\dal \fbyf{f(g[C])}{\sigma_{\bet\gam}(r)} = f'(g[C])\dal
\fbyf{g[C]}{\sigma_{\bet\gam}(r)}
\ee
In other words, this operator acts as a first order derivative on the loop
functional with finite area derivative (so called Stokes type functional).
Then, the above term does not appear.

The Action functional $ S[C] $ describes the distribution of the large
scale vorticity structures, and hence it should not depend of viscosity.
In terms of the above connected correlation functions of the circulation
this corresponds to the limit, when the effective Reynolds number
$\frac{\Gamma_C[v]}{\nu}$ goes to infinity, but the sum of the divergent
series tends to the finite limit. According to the standard picture of
turbulence, the large scale vorticity structures depend upon the energy
pumping, rather than the energy dissipation.

It is understood that  both time $ t $ and the loop
size\footnote{As a measure of the loop size one may take the square root
of the minimal area inside the
loop.} $ |C| $ should be greater then the viscous scales
\begin{equation}
t \gg t_0 = \nu^{\frac{1}{2}}{\cal E}^{-\frac{1}{2}} \\;\;
|C| \gg r_0 = \nu^{\frac{3}{4}} {\cal E}^{-\frac{1}{4}}
\end{equation}
where $ {\cal E } $ is the energy dissipation rate.

It is defined from the energy balance equation
\be
0 = \d_t\VEV{\oh\,\val^2}= \nu \VEV{\val\d^2 \val} +\VEV{f_{\alp}\val}
\ee
which can be transformed to
\be
 \oq \nu \VEV{\omega_{\alp\bet}^2} = 3 F(0)
\ee
The left side represents the energy, dissipated at small scale due to
viscosity, and the right side - the energy pumped in from the large scales
due to the random forces. Their common value is $\et$.

We see, that constant $F(r-r')$, i.e., $\tilde{F}(k)\propto \delta(k)$ is
sufficient to  provide the necessary energy pumping. However, such forcing
does not produce vorticity, which we readily see in our equation. The
contribution from this constant part to the potential in our loop equation
drops out (this is a  closed loop integral of total derivative). This is
important, because this term would have the wrong order of magnitude in
the turbulent limit - it would grow as the Reynolds number.

Dropping this term, we arrive at remarkably simple and universal
functional equation
\be
\dot{S}[C] =  \oint_{C} dr_{\alpha}
\int d^3 r'\frac{r'_{\gamma}-r_{\gam}}{4 \pi |r-r'|^3}
\frac{\delta S}{\delta \sigma_{\beta \alpha}(r)}
\frac{\delta S}{\delta \sigma_{\beta \gamma}(r')}
\label{KIN}
\ee
The stationary solution of this equation describes the steady
distribution of the circulation in the strong turbulence. Note, that the
stationary solutions  come in pairs $ \pm S$. The sign should be chosen
so, that $ \Im S > 0 $, to provide the inequality $ |\Psi| <1$.

\section{Scaling law}

The `Hamilton-Jacobi' equation  without the potential term (\ref{KIN})
allows the family of the scaling solutions
\begin{equation}
S[C] = t^{2 \kappa -1}\phi \left[\frac{C}{t^{\kappa}} \right]
\end{equation}
with arbitrary index $ \kappa $. The scaling function satisfies the
equation
\begin{equation}
(2 \kappa -1 ) \phi[C]  - \kappa \oint_{C} dr_{\alpha}
\frac{\delta \phi[C]}{\delta \sigma_{\beta \alpha}(r)}r_{\beta} =
\oint_{C} dr_{\alpha}
\int d^3 r'\frac{r'_{\gamma}-r_{\gam}}{4 \pi |r-r'|^3}
\frac{\delta \phi[C]}{\delta \sigma_{\beta \alpha}(r)}
\frac{\delta \phi[C]}{\delta \sigma_{\beta \gamma}(r')}
\end{equation}
The left side here was computed, using the chain rule differentiation of
functional.

Asymptotically, at large time, we expect the fixed point, which is the
homogeneous functional
\begin{equation}
S_{\infty}[C] = |C|^{2- \frac{1}{\kappa}} f \left[\frac{C}{|C|} \right]
\end{equation}
zeroing the right side of our `kinetic' functional equation
\begin{equation}
0=\oint_{C} dr_{\alpha}
\int d^3 r'\frac{r'_{\gamma}-r_{\gam}}{4 \pi |r-r'|^3}
\frac{\delta S_{\infty}[C]}{\delta \sigma_{\beta \alpha}(r)}
\frac{\delta S_{\infty}[C]}{\delta \sigma_{\beta \gamma}(r')}
\end{equation}

The Kolmogorov scaling \cite{Kolm41} would correspond to
\begin{equation}
\kappa = \frac{3}{2}
\end{equation}
in which case one can express the $ S $ functional in terms of $ {
\cal E } $
\begin{equation}
S[C] = {\cal E} t^2 \phi \left[\frac{C}{\sqrt{{\cal E}t^3}} \right]
\end{equation}

One can easily rephrase the Kolmogorov arguments in the loop space.
The relation between the energy dissipation rate and the velocity
correlator reads
\begin{equation}
{\cal E } = \left \langle v_{\alpha}(r_0) v_{\beta}(0) \partial_{\beta}
v_{\alpha}(0) \right \rangle
\end{equation}
where the point splitting at the viscous scale $r_0$ is introduced. Such
splitting is necessary to avoid the viscosity effects; without the
splitting the average would formally reduce to the total derivative
and vanish.

Instead of the point splitting one may introduce the finite loop of
the viscous scale $ |C| \sim r_0 $, and compute this correlator in
presence of such
loop. This reduces to the WKB estimates
\begin{equation}
\omega_{\alpha \beta}(r) \rightarrow
\frac{\delta S[C]}{\delta \sigma_{\alpha \beta}(r)} \\;\;
v_{\alpha}(r) = \int d^3 r'\frac{r'_{\gamma}-r_{\gam}}{4 \pi |r-r'|^3}
\omega_{\alpha \gamma}(r')
\end{equation}

Using the generic scaling law for $ S $ we find
\begin{equation}
\omega \sim r_0^{- \frac{1}{\kappa}}\\;\;
v \sim r_0^{1-  \frac{1}{ \kappa}}\\;\;
{\cal E} \sim r_0^{2 - \frac{3}{\kappa}}
\end{equation}

We see, that the energy dissipation rate would stay finite in the limit of
the vanishing viscous scale only for the Kolmogorov value of the index.
This argument looks rather cheap, but I think it is basically
correct. The constant value of the energy dissipation rate in the limit
of vanishing viscosity arises as the quantum anomaly in the field
theory, through the finite limit of the point splitting in the
correspondent energy current.\footnote{I am grateful to A.~Polyakov and
E.~Siggia for inspiring comments on this subject.}

There is another version of this argument, which I like better. The
dynamics of  Euler fluid in infinite system would not exist, for the
non-Kolmogorov scaling. The extra powers of loop size would have to enter
with the size $L$ of the whole system, like
$\left(\frac{|C|}{L}\right)^{\eps} $. So, in the regime with finite energy
pumping rate $\et$ the infinite Euler system can exist only for the
Kolmogorov index. This must be the essence of the original Kolmogorov
reasoning \ct{Kolm41}.

The problem is that nobody proved that such limit exists, though. Within
the usual framework, based on the velocity correlation functions, one has
to prove, that the infrared divergencies, caused by the sweep, all cancel
for the observables. Within our framework these problems disappear, as we
shall see later.

As for the correlation functions in inertial range, unfortunately
those cannot be computed in the WKB approximation, since they involve the
contour shrinking to a double line, with vanishing area inside. Still,
most of the physics can be understood in loop terms, without these
correlation functions. The large scale behavior of the loop functional
reflects the statistics of the large vorticity structures, encircled by
the loop.

\section{Loop Equation for the Circulation PDF}

The loop field could serve as the generating function for the PDF $P_C(\Gamma)
$ for the circulation. The Fourier integral
\be
P_C(\Gamma) = \int_{-\8}^{\8} \frac{d g}{2 \pi \nu}
\EXP{ \frac{\i g }{\nu} \left(\C \ral \val(r)  - \Gamma \right)}
\ee
can be analyzed in the same way as the loop field before. The only difference
is that  the factors of $ g $ appear in front of various terms. These factors
can be replaced by
\be
g \ra \i \nu \pp{\Gamma}
\ee
acting on $P_C(\Gamma) $.

As a result we find
\bea
 \pp{\Gamma}\dot{P}_C(\Gamma) = -\oint_{C} dr_{\alpha}
\int d^3 r'\frac{r'_{\gamma}-r_{\gam}}{4 \pi |r-r'|^3}
\frac{\delta^2 P_C(\Gamma)}{\delta \sigma_{\beta \alpha}(r)
\delta \sigma_{\beta \gamma}(r')} \br
 +\nu \pp{\Gamma}
\oint_{C} dr_{\alpha}\partial_{\beta}
     \frac{\delta P_C(\Gamma)}{\delta \sigma_{\beta \alpha}(r)}
- U[C] \frac{\d^3 P_C(\Gamma)}{\d \Gamma^3}
\label{PDF}
\eea
All the imaginary units disappear, as they should. As for the viscosity and
forcing, these terms can be neglected in inertial range in the same way as
before. The only new thing is that one has to assume that $ \Gamma \gg \nu $ in
inertial range in addition to above assumptions about the size of the loop.

In absence of these terms there are no dimensional parameters so that the
following scaling laws hold (with the same index $ \kappa $ as before)
\be
P_C(\Gamma) = t^{2 \kappa-1} F\left[\frac{C}{t^{\kappa}},\frac{\Gamma}{t^{2
\kappa-1}} \right]
\ee
The factor $ t^{2 \kappa-1} $ came from the normalization of probability
density. Note, that this is more general law than before. Here we do not have
to use the WKB approximation for the PDF. In other words, the whole PDF rather
than just its decay  at large $ \Gamma $ satisfies the scaling law.

The steady distribution would have the form of
\be
P_C(\Gamma) \ra \inv{\Gamma}\Phi\left[\frac{C}{\Gamma^{\frac{\kappa}{2
\kappa-1}}} \right]
\ee
where the scaling functional $ \Phi $ satisfies the homogeneous equation
\be
\oint_{C} dr_{\alpha}
\int d^3 r'\frac{r'_{\gamma}-r_{\gam}}{4 \pi |r-r'|^3}
\frac{\delta^2 \Phi[C]}{\delta \sigma_{\beta \alpha}(r)
\delta \sigma_{\beta \gamma}(r')} =0
\ee
with the normalization condition
\be
1=\int_{-\8}^{\8} \frac{d \Gamma}{ \Gamma}
\Phi\left[\frac{C}{\Gamma^{\frac{\kappa}{2 \kappa-1}}} \right]
\ee

In principle, there could be different scaling functions for positive and
negative $ \Gamma $, rather than just absolute value $ |\Gamma| $ prescription.
This would correspond to above mentioned violation of the time reversal
symmetry.
However, as we mentioned above, there is no exact relation which would
eliminate the symmetric solution.

The Kolmogorov triple correlation function vanishes for vorticities (see
Appendix I), so that there is no restriction on the asymmetric part of the
circulation PDF. Nevertheless, the Kolmogorov scaling $ \kappa = \frac{3}{2} $
seems to me the most likely possibility, by the reasons discussed in the
previous section.

The homogeneous loop equation requires some boundary conditions at large loops,
to provide a meaningful solution. The asymptotic decrease of PDF
\be
P_C(\Gamma) \sim \EXP{- Q\left[\frac{C}{\Gamma^{\frac{\kappa}{2 \kappa-1}}}
\right]}, Q \ra \8
\ee
would lead to the same WKB equation as before
\be
\oint_{C} dr_{\alpha}
\int d^3 r'\frac{r'_{\gamma}-r_{\gam}}{4 \pi |r-r'|^3}
\frac{\delta Q[C]}{\delta \sigma_{\beta \alpha}(r)}
\frac{\delta Q[C]}{\delta \sigma_{\beta \gamma}(r')} =0
\ee
We are studying this equation  in the next section.

\section{Tensor Area law}

The Wilson loop in QCD decreases as exponential of the minimal area,
encircled by the loop, leading to the quark confinement. What is the
similar asymptotic law in turbulence? The physical mechanisms leading to
the area law in QCD are absent here. Moreover, there is no guarantee, that
$\Psi[C]$ always decreases with the size of the loop.

This makes it possible to look for the simple Anzatz, which was not
acceptable in QCD, namely
\be
S[C] = s\left(\Sigma_{\mu\nu}^C\right)
\ee
where
\be
\Sigma_{\mu\nu}^C= \oint_C r_{\mu} d r_{\nu}
\ee
is the tensor area encircled by the loop $C$. The difference between this
area and the scalar area is the positivity property. The scalar area
vanishes only for the loop which can be contracted to a point by removal
of all the backtracking. As for the tensor area, it vanishes, for example,
for the $8$ shaped loop, with opposite orientation of petals.

Thus, there are some large contours with vanishing tensor area, for which
there would be no decrease of the $\Psi$ functional.
In QCD the Wilson loops must always decrease at large distances, due to
the finite mass gap. Here, the large scale correlations are known to
exist, and play the central role in the turbulent flow. So, I  see no
convincing arguments to reject the tensor area Anzatz.

This Anzatz in QCD not only was unphysical, it failed to reproduce the
correct short-distance singularities in the loop equation. In turbulence,
there are no such singularities. Instead, there are the  large-distance
singularities, which all should cancel in the loop equation.

It turns out, that  for this Anzatz the (turbulent limit of the) loop
equation is satisfied automatically, without any further restrictions.
Let us verify this important property. The first area derivative yields
\be
\omega_{\mu\nu}^C(r)=\fbyf{S}{\sigma_{\mu\nu}(r)} = 2\pbyp{s}{
\Sigma_{\mu\nu}^C}
\ee
The factor of $2$ comes from the second term in the variation
\be
\fbyf{\Sigma^C_{\alp\bet}}{\sigma_{\mu\nu}(r)}=
\del_{\alp\mu}\del_{\bet\nu}-\del_{\alp\nu}\del_{\bet\mu}
\ee
Note, that the right side does not depend on $r$. Moreover, you can shift
$r$ aside from the base loop $C$, with proper wires inserted. The area
derivative would not change, as the contribution of wires drops.

This implies, that the corresponding vorticity $\omega_{\mu\nu}^C(r) $ is
space independent, it only depends upon the loop itself. The velocity can
be reconstructed from vorticity up to irrelevant constant sterms
\be
\vbe^C(r) = \oh\,\ral\,\omega_{\alp\bet}^C
\ee
This can be formally obtained from the above integral representation
\be
\vbe^C(r) =\int d^3 r'\frac{\ral-r'_{\alp}}{4 \pi
|r-r'|^3}\omega_{\alp\bet}^C
\label{INTG}
\ee
as a residue from the infinite sphere $ R = |r'| \ra \8$. One may insert
the regularizing factor $ |r'|^{-\epsilon}$ in $\omega$, compute the
convolution integral in Fourier space and check that in the limit $
\epsilon  \ra 0^+$ the above linear term arises. So, one can use the above
form of the loop equation, with the analytic regularization prescription.

Now, the $v\,\omega$ term in the loop equation reads
\be
\oint _C d r_{\gam} \,\vbe^C(r)\,\omega_{\bet\gam}^C \propto
\Sigma_{\gam\alp}^C\,\omega_{\alp\bet}^C\,\omega_{\bet\gam}^C
\ee
This tensor trace vanishes, because the first tensor is antisymmetric, and
the product of the last two antisymmetric tensors is symmetric with
respect to $\alp\gam$.

So, the positive and negative terms cancel each other in our loop
equation, like the "income" and "outcome" terms in the usual kinetic
equation. We see, that there is an equilibrium  in our loop space kinetics.

{}From the point of view of the notorious infrared divergencies in
turbulence, the above calculation explicitly demonstrates how they cancel.
By naive dimensional counting these terms were linearly divergent. The
space isotropy lowered this to logarithmic divergency in \rf{INTG}, which
reduced to finite terms at closer inspection. Then, the explicit form of
these terms was such, that they all cancelled.

This cancellation originates from the angular momentum conservation in
fluid mechanics.  The large loop $C$  creates the macroscopic eddy with
constant vorticity  $\omega_{\alp\bet}^C$ and linear velocity $ v^C(r)
\propto r$. This is a well known static solution of the \NS equation. The
eddy is conserved due to the angular momentum conservation.The only
nontrivial thing  is the functional dependence of the eddy vorticity upon
the shape and size of the loop $C$.  This is a function of the tensor area
$\Sigma_{\mu\nu}^C$, rather than a general functional of the loop.

Combining this Anzatz with the space isotropy and the Kolmogorov scaling
law, we arrive at the tensor area law
\be
\Psi[C] \propto \EXP{-
B\,\left(\frac{\et}{\nu^3}\left(\Sigma^C_{\alp\bet}\right)^2\right)^{\ot} }
\label{AREA}
\ee
The universal constant $B$ here must be real, in virtue of the loop
orientation symmetry. When the orientation is reversed $C(\theta) \ra
C(2\pi-\theta)$, the loop integral changes sign, but its square, which
enters here, stays invariant. Therefore, the constant in front must be
real. The time reversal tells the same, since {\em both} viscosity $\nu$
and the energy dissipation rate $\et$ are time-odd.  Therefore, the ratio
$\frac{\et}{\nu^3}$ is time-even, hence it must enter $\Psi[C]$ with the
real coefficient. Clearly, this coefficient $B$ must be positive, since $
\left|\Psi[C] \right|<1$.

Note, however, that we did not prove this law. The absence of decay for large
twisted loops with zero tensor area is suspicious. Also, the physics seems to
be different from what we expect in turbulence. The uniform vorticity, even a
random one, as in this solution, contrasts the observed intermittent
distribution. Besides, there clearly must be corrections to the asymptotic law,
whereas the tensor area law is {\em exact}. This is far too simple. We
discussed this unphysical solution mostly as a test of the loop technology.

\section{Scalar Area law}

Let us now study the scalar area law, which is a valid Anzatz for the
asymptotic decay of the circulation PDF.
The set of equations for the minimal surface  is summarized in Appendix A.
All we need here is the following representation
\be
A \ra \inv{2L^2_{\Gamma}}\,
\int \int d \sigma_{\mu\nu}(x) d \sigma_{\mu\nu}(y)
\EXP{-\pi\frac{(x-y)^2}{L^2_{\Gamma}}}
\ee
where $ L_{\Gamma} =|\Gamma|^{\tq} \et^{-\oq} $.
The distance $ (x-y)^2$ is measured in 3-space and integration goes
along the minimal surface. It is implied that  its size is much larger than
$ L_{\Gamma} $.

In this limit the integration over, say, $ y $ can be performed along the
local tangent plane at $ x $ in small vicinity $ y-x \sim L_{\Gamma} $ ,
after which the factors of $ L_{\Gamma} $ cancel. We are left then with the
ordinary scalar area integral
\be
 A \ra \oh \int d\sigma_{\mu\nu}(x)
 d \sigma_{\mu\nu}(y) \delta^2(x-y) \ra \int d^2x \sqrt{g}
\ee

In the previous, regularized form the area represents so called Stokes
functional\ct{Mig83}, which can be substituted into the loop equation.
The area derivative of the area reads
\be
\fbyf{A}{\sigma_{\mu\nu}(x)} =
\inv{L^2_{\Gamma}}\, \int  d \sigma_{\mu\nu}(y)
\EXP{-\pi\frac{(x-y)^2}{L^2_{\Gamma}}}
\ee
In the local limit this reduces to the tangent tensor
\be
\fbyf{A}{\sigma_{\mu\nu}(x)} \ra \int  d \sigma_{\mu\nu}(y) \delta^2(x-y)
= t_{\mu\nu}(x)
\ee
It is implied that the point $x$ approaches the contour from inside the
surface, so that the tangent tensor is well defined
\be
t_{\mu\nu}(x) = t_{\mu}n_{\nu} - t_{\nu} n_{\mu}
\ee
Here $ t_{\mu}$ is the local tangent vector of the loop, and $ n_{\nu} $
is the inside normal to the loop along the surface.

The second area derivative of the regularized area in this limit is just
the exponential
\be
\frac{\delta^2
A}{\delta\sigma_{\alpha\beta}(x)\delta\sigma_{\gamma\delta}(y)}=
\inv{L^2_{\Gamma}}\, \EXP{-\pi\frac{(x-y)^2}{L^2_{\Gamma}}}
\ee
Should we look for the higher terms of the asymptotic expansion at large
area we would have to take into account the shape of the minimal
surface, but in the thermodynamical limit we could neglect the curvature of
the loop and use the planar disk.

Let us use the general WKB form of PDF
\be
P_C(\Gamma) = \inv{\Gamma}
\EXP{-Q\left(\frac{A}{t^{2\kappa}},\frac{\Gamma}{t^{2 \kappa-1}} \right)}
\ee
We shall skip the arguments of effective action $ Q $.
We find on the left side of the loop equation
\be
\d_t Q \d_{\Gamma} Q - \d_t \d_{\Gamma} Q
\ee
On the right side we find  the following integrand
\be
\left(\left(\d_{A}Q\right)^2-\d^2_{A}Q\right)
\,\fbyf{A}{\sigma_{\alpha\beta}(r)}
\fbyf{A}{\sigma_{\gamma\beta}(r')} -
\d_{A}Q \,\frac{\delta^2 A}{\delta\sigma_{\alpha\beta}(r)
\delta\sigma_{\gamma\beta}(r')}
\ee

The last term drops after the  $r' $ integration in virtue of symmetry.
The leading terms in the WKB approximation on both sides are those with
the first derivatives. We find
\be
\d_t Q \d_{\Gamma}Q =
\left(\d_{A}Q\right)^2 \C \ral \fbyf{A}{\sigma_{\alpha\beta}(r)}
\int d^3 r' \frac{\rga-r'_{\gamma}}{4\pi|r-r'|^3}\,
\fbyf{A}{\sigma_{\gamma\beta}(r')}
\ee

In the last integral we substitute above explicit form of the
area derivatives and perform the $ d^3r' $ integration first. In the
thermodynamical limit only the small vicinity $ r'-y \sim L_{\Gamma} $
contributes, and we find
\be
\int d^3 r' \frac{\rga-r'_{\gamma}}{4\pi|r-r'|^3}\,
\fbyf{A}{\sigma_{\gamma\beta}(r')} \ra
L_{\Gamma}^2 \,\int d \sigma_{\gamma\beta}(y)
\frac{\rga-y_{\gamma}}{4\pi|r-y|^3}
\ee

This integral logarithmically diverges. We compute it with
the logarithmic accuracy with the following result
\be
\int d \sigma_{\gamma\beta}(y)
\frac{\rga-y_{\gamma}}{4\pi|r-y|^3}
\propto \frac{t_{\beta}}{ \pi} \ln \frac{L^2_{\Gamma}}{A}
\ee
The meaning of this integral is the average velocity in the WKB
approximation. This velocity is tangent to the loop, up to the next
correction terms at large area.

Now, the emerging loop integral vanishes due to symmetry
\be
\C \ral t_{\beta} t_{\alpha\beta} =0
\ee
as the line element $ d \ral $ is directed along the tangent vector $
t_{\alpha} $, and the tangent tensor $ t_{\alpha\beta} $ is antisymmetric.
Similar mechanism was used in the tensor area solution, only there the
cancellations emerged at the global level, after the closed loop
integration. Here the right side of the loop equation vanishes locally,
at every point of the loop. Anyway, we see, that the scalar area indeed
represents the steady solution of the loop equation in the leading WKB
approximation.

It might be instructive to compare this solution with another known exact
solution of the Euler dynamics, namely the Gibbs solution
\be
P[v] = \EXP{-\beta \int d^3 r \oh \val^2 }
\ee
For the loop functional it reads
\be
\Y = \EXP{-\frac{\gamma^2}{2\beta}\C \ral \C r'_{\beta} \delta^3(r-r')}
\ee
The integral  diverges, and it corresponds to the perimeter law
\be
\C \ral \C r'_{\beta} \delta^3(r-r') \ra r_0^{-2} \oint_C |dr|
\ee
where $ r_0 $ is a small distance cutoff. For the PDF it yields
\be
P(\Gamma) \propto \EXP{-\frac{\Gamma^2\beta r_0^2}{2 \oint_C |dr|}}
\ee

When the Gibbs solution is substituted into the loop equation, we observe
the same thing. Average velocity is tangent to the loop, which leads to
vanishing integrand in the loop equation. The difference is that in our
case this is true only asymptotically, there are next corrections.

The shape of the function $ Q $ is not fixed by this equation in the leading
WKB approximation.  In a scale invariant theory it is natural to expect the
power law
\be
Q\left(\frac{A}{t^{2\kappa}},\frac{\Gamma}{t^{2 \kappa-1}} \right) \ra
\mbox{const } \left(\Gamma^{2\kappa} A^{1-2\kappa}\right)^{\mu}
\label{MuLaw}
\ee
There is one more arbitrary index $ \mu$ involved. Even for the Kolmogorov law
$ \kappa = \frac{3}{2} $ the $ \Gamma $ dependence remains unknown.

\section{Discussion}

So, we found  two asymptotic solutions of the loop equation in the
turbulent limit, not counting the Gibbs solution. It remains to be seen, which
one (if any) is realized in turbulent flows. The tensor area solution is
mathematically cleaner, but its physical meaning contradicts the intermittency
paradigm. It corresponds to the uniform vorticity with random magnitude and
random direction, rather that the regions of high vorticity interlaced with
regions of low vorticity, observed in the turbulent flows.

The recent numerical experiments\cite{Umeki} favor the scalar area rather than
the tensor one. Also, the Kolmogorov scaling was observed in these experiments.
The Reynolds number was only $ \sim 100 $ which was too small to make any
conclusions. We have to wait for the experiments ( real or numerical ) with the
Reynolds numbers few orders of magnitude larger.

The scalar area is less trivial than the tensor one. The minimal area as a
functional of the loop cannot be represented as any explicit contour integral
of the Stokes type, therefore it corresponds to infinite number of higher
correlation functions present. Moreover, there could be several minimal
surfaces for the same loop, as the equations for the minimal surface are
nonlinear. Clearly, the one with the least area should be taken.

The natural generalization of this solution  is the string Anzatz where the sum
over all surfaces bounded by the loop is taken
\be
P_C(\Gamma) = \sum_{S: \d
S=C}\EXP{-Q\left(\frac{A}{t^{2\kappa}},\frac{\Gamma}{t^{2 \kappa-1}} \right)}
\ee
At large loop the minimal $ Q $ terms will remain. The extremum condition
\be
\delta Q = \frac{\d Q}{\d A } \delta A =0
\ee
will be satisfied for the minimal surface.

However, the  sum over random surfaces is not well defined. The recent
studies\ct{QG} indicate that the typical closed surfaces degenerate to branched
polymers. For the surface bounded by a fixed loop this cannot happen, of
course. Still nobody knows how to compute such sums. The loop equation in
principle allows to systematically  compute the corrections  to the area law as
the WKB expansion.

The WKB solution is incomplete so far. The leading term in the loop equation is
annihilated by arbitrary function of the area (scalar or tensor). The similar
ambiguity was present in the Gibbs solution, where arbitrary function of the
Hamiltonian satisfied the Liouville equation for the velocity PDF.  In that
case the ambiguity was removed by extra requirement of  thermodynamic limit:
only the exponential of the hamiltonian would agree with the factorization of
the PDF for two remote parts of the system.

What could be a similar requirement here? The area of the minimal surface
represents the effective volume of the system at large loop. The circulation
can be written as a surface integral of vorticity,  which makes the circulation
an extensive variable at this surface.  The average vorticity
\be
\bar{\omega} = \frac{\Gamma}{A}
\ee
represents an intensive quantity.  The thermodynamic limit would then
correspond to
\be
Q = A q(\bar{\omega})
\ee

Comparing this with the previous formula for $ Q $ \rf{MuLaw} we conclude that
$ \mu=1$. In this case
\be
Q =\mbox{const }A  \bar{\omega}^{2\kappa}
\ee
In principle, there could be two different laws for positive and negative $
\Gamma $, due to violation of the time reversal invariance
\be
Q \ra q_{\pm} A |\bar{\omega}|^{2\kappa}
\ee

Another line of argument  might start with an assumption of decorrelated
average vorticity  $ \omega_i $ at various parts $ S_i $ of the area $ A_0 $ of
the minimal surface. The net circulation, adding up from the large number $ n
\sim \frac{A}{A_0} \gg 1 $ of independent random terms $ \omega_i A_0 $ would
be a gaussian variable as a consequence of the law of large numbers. We would
have then
\be
Q \sim \frac{\Gamma^2}{n A_0^2 \omega_i^2} = \frac{\Gamma^2 }{A A_0\omega_i^2}
\ee
This  would agree with the previous estimate at
\be
 \mu = \inv{\kappa},  A_0\omega_i^2\sim A^{1- \inv{\kappa}}
\ee
so that
\be
Q \ra \mbox{const } \Gamma^2 A^{\inv{\kappa}-2}
\ee
The natural assumption here would be that the vorticity variance $ \omega_i$
does not scale with the area $ A $, so that
\be
A_0 \sim A^{1- \inv{\kappa}}
\ee

The Gaussian behavior (with $ \kappa = \frac{3}{2}  $ ) was observed in
numerical experiments \ct{Umeki}, but the Reynolds number was too low to make
conclusions at this point. There could be  a scaling function
\be
Q = q\left(\Gamma^{2} A^{\inv{\kappa}-2}\right)
\ee
which starts  linearly and then grows as  a power, say, $ q(x) = (1+ a
\,x)^{\kappa} $.  I suggest that this function should be studied in real and
numerical experiments. This would teach us something new about turbulence.

\section{Acknowledgments}

I am grateful to V.~Borue, I.~Goldhirsh, D.~McLaughlin, A.~Polyakov and
V.~Yakhot for stimulating discussions .

\newpage

\appendix

\section{Loop Expansion}

Let us outline the method of direct iterations of the loop equation. The
full description of the method can be found in \cite{Mig83}. The basic
idea is to use the following representation of the loop functional
\be
\Psi[C] = 1+\sum_{n=2}^{\8} \inv{n} \left\{\oint_C dr_1^{\alp_1} \dots
\oint_C dr_n^{\alp_n}\right\}_{\mbox{cyclic}} W^n_{\alp_1\dots
\alp_n}\left(r_1,\dots r_n\right)
\label{STOKES}
\ee

This representation is valid for every translation invariant functional
with finite area derivatives (so called Stokes type functional). The
coefficient functions $W$ can be related to these area derivatives. The
normalization $\Psi[0]=1 $ for the shrunk loop is implied.

In general case the integration points $r_1,\dots  r_n$ in \rf{STOKES} are
cyclicly ordered around the loop $C$. The coefficient functions can be
assumed cyclicly symmetric without loss of generality. However,  in case
of fluid dynamics, we are dealing with so called abelian Stokes
functional. These functionals are characterized by completely symmetric
coefficient functions, in which case the ordering of points can be
removed, at expense of the extra symmetry  factor in denominator
\be
\Psi[C] = 1+\sum_{n=2}^{\8} \inv{n!} \oint_C dr_1^{\alp_1} \dots  \oint_C
dr_n^{\alp_n} W^n_{\alp_1\dots \alp_n}\left(r_1,\dots r_n\right)
\label{ABEL}
\ee
The incompressibility conditions
\be
\d_{\alp_k}W^n_{\alp_1\dots \alp_n}\left(r_1,\dots r_n\right)=0
\label{divv}
\ee
does not impose any further restrictions, because of the gauge invariance
of the loop functionals. This invariance (nothing to do with the symmetry
of dynamical equations!) follows from the fact, that the closed loop
integral of any total derivative  vanishes. So, the coefficient functions
are defined modulo such derivative terms. In effect this means, that one
may relax the incompressibility constraints \rf{divv}, without changing
the loop functional.

To avoid confusion, let us note, that the physical incompressibility
constrains are not neglected. They are, in fact, present in the loop
equation, where we used the integral representation for the velocity in
terms of vorticity. Still, the longitudinal parts of $W$ drop in the loop
integrals.

The loop calculus for the abelian Stokes functional is especially simple.
The area derivative corresponds to removal of one loop integration, and
differentiation of the corresponding coefficient function
\be
\fbyf{\Psi[C]}{\sigma_{\mu\nu}(r)} = \sum_{n=1}^{\8} \inv{n!} \oint_C
dr_1^{\alp_1} \dots  \oint_C dr_n^{\alp_n}
\hat{V}_{\mu\nu}^{\alp}W^{n+1}_{\alp,\alp_1\dots \alp_n}\left(r,r_1,\dots
r_n\right)
\label{ABEL'}
\ee
where
\be
\hat{V}_{\mu\nu}^{\alp} \equiv
\d_{\mu}\delta_{\nu\alp}-\d_{\nu}\delta_{\mu\alp}
\ee
In the nonabelian case, there would also be the contact terms, with $W$ at
coinciding points, coming  from the cyclic ordering \ct{Mig83}. In abelian
case these terms are absent, since $W$ is completely symmetric.

As a next step, let us compute the local kinetic term
\be
\hat{L} \Psi[C]  \equiv \oint_C d r_{\nu}
\d_{\mu}\fbyf{\Psi[C]}{\sigma_{\mu\nu}(r)}
\ee
Using above formula for the loop derivative, we find
\be
\hat{L} \Psi[C]  = \sum_{n=1}^{\8} \inv{n!}  \oint_C dr^{\alp}\oint_C
dr_1^{\alp_1} \dots  \oint_C dr_n^{\alp_n} \d^2W^{n+1}_{\alp,\alp_1\dots
\alp_n}\left(r,r_1,\dots r_n\right)
\label{L}
\ee
The net result is the second derivative of $W$ with respect to one
variable. Note, that the second term in $\hat{V}_{\mu\nu}^{\alp}$ dropped,
as the total derivative in the closed loop integral.

As for the nonlocal kinetic term, it involves the second area derivative
off the loop, at the point $r'$, integrated over $r'$ with the
corresponding Green's function.  Each area derivative involves the same
operator $\hat{V}$, acting on the coefficient function. Again, the abelian
Stokes functional simplifies the general framework of the loop calculus.
The contribution of the wires cancels here, and the ordering does not
matter, so that
\be
\frac{\delta^2\Psi[C]}
{\delta\sigma_{\mu\nu}(r)\delta\sigma_{\mu'\nu'}(r')} = \sum_{n=0}^{\8}
\inv{n!} \oint_C dr_1^{\alp_1} \dots  \oint_C dr_n^{\alp_n}
\hat{V}_{\mu\nu}^{\alp}\hat{V'}_{\mu'\nu'}^{\alp'}W^{n+2}_{\alp,\alp',\alp_1\dots
\alp_n}\left(r,r',r_1,\dots r_n\right)
\ee

Using these relations, we can write the steady state loop equation as
follows

\pct{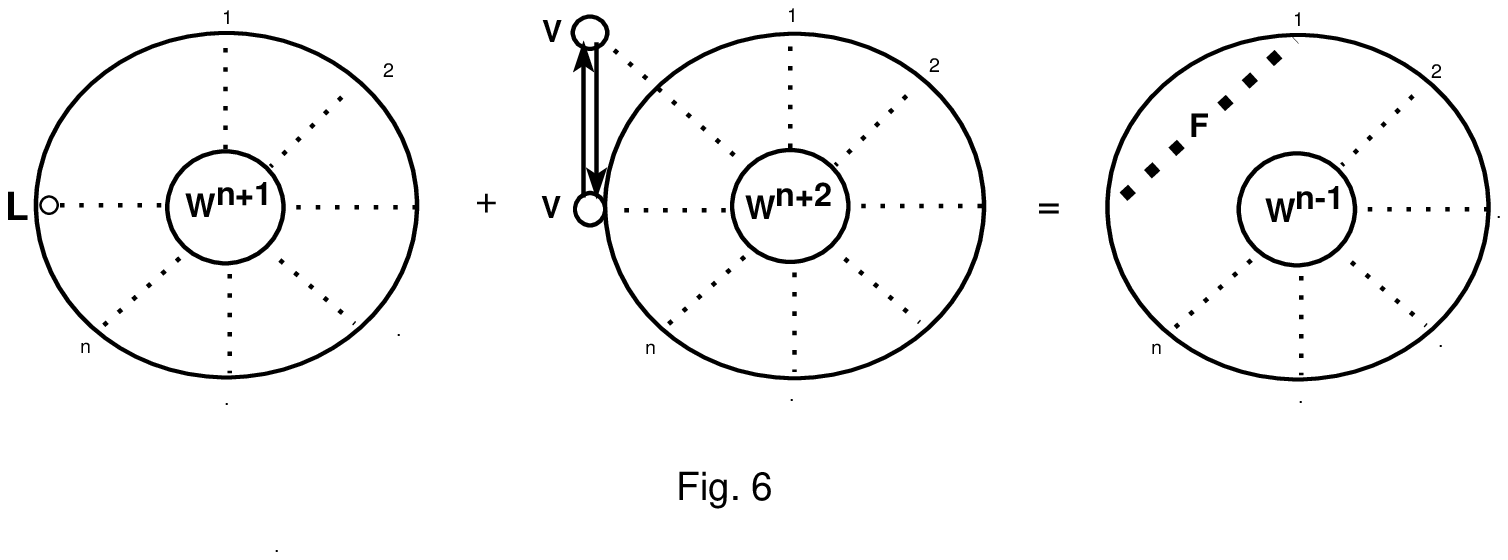}
Here the light dotted lines symbolize the arguments $\alp_k, r_k$ of $W$,
the big circle denotes the loop $C$, the tiny circles stand for the loop
derivatives, and the pair of lines with the arrow denote the Green's
function. The sum over the tensor indexes and the loop integrations over
$r_k$ are implied.

The first term is the local kinetic term, the second one is the nonlocal
kinetic term, and the right side is the potential term in the loop
equation. The heavy dotted line in this term stands for the correlation
function $F$ of the random forces. Note that this term is an abelian
Stokes functional as well.

The iterations go in the potential term, starting with $\Psi[C]=1$. In the
next approximation, only the two loop correction
$W^2_{\alp_1\alp_2}(r_1,r_2)$ is present.  Comparing the terms, we note,
that nonlocal kinetic term reduces to the total derivatives due to the
space symmetry (in the usual terms it would be $ \VEV{v\omega}$ at
coinciding arguments), so we are left with the local one.

This yields the equation
\be
\nu^3 \d^2 W^2_{\alp\bet}(r-r') = F(r-r')\delta_{\alp\bet}
\ee
modulo derivative terms.  The solution is trivial in Fourier space
\be
 W^2_{\alp\bet}(r-r') = -\int \frac{d^3 k}{(2\pi)^3} \EXP{\i k (r-r')}
\delta_{\alp\bet} \frac{\tilde{F}(k)}{\nu^3 k^2}
\ee
Note, that we did not use the transverse tensor
\be
P_{\alp\bet}(k) = \delta_{\alp\bet}- \frac{k_{\alp} k_{\bet}}{k^2}
\ee
Though such tensor is present in the physical velocity correlation, here
we may use $\del_{\alp\bet}$ instead, as the longitudinal terms drop in
the loop integral. This is analogous to the Feynman gauge in QED. The
correct correlator corresponds to the Landau gauge.

The potential term generates the  four point correlation $ F \,W^2$. which
agrees with the disconnected term in the $W^4$ on the left side
\bea
W^4_{\alp_1\alp_2\alp_3\alp_4}\left(r_1,r_2,r_3,r_4\right) \ra
W^2_{\alp_1\alp_2}\left(r_1-r_2\right)W^2_{\alp_3\alp_4}\left(r_3-r_4\right)
+ \br
W^2_{\alp_1\alp_3}\left(r_1-r_3\right)W^2_{\alp_2\alp_4}\left(r_2-r_4\right)
+
W^2_{\alp_1\alp_4}\left(r_1-r_4\right)W^2_{\alp_2\alp_3}\left(r_2-r_3\right)
\eea
In the same order of the loop expansion, the three point function will
show up. The corresponding terms in kinetic part must cancel among
themselves, as the potential term does not contribute. The local kinetic
term yields the loop integrals of $ \d^2 W^3 $, whereas the  nonlocal one
yields $\hat{V}W^2 \,\hat{V'}W^2$, integrated over $d^3 r'$ with the
Greens's function $ \frac{(r-r')}{4\pi |r-r'|^3}$. The equation has the
structure

\pct{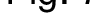}

Now it is clear, that the solution of this equation for $W^3$ would be the
same three point correlator, which one could obtain (much easier!) by
direct iterations of the \NS equation.

The purpose of this painful exercise was not to give one more method of
developing the expansion in powers of the random force. We rather
verified  that the loop equations are capable of producing the same
results, as the ordinary chain of the equations for the correlation
functions.

In above arguments, it was important, that the loop functional belonged to
the class of the abelian Stokes functionals. Let us check that our tensor
area Anzatz
\be
\Sigma^C_{\alp\bet}=\oint_C \ral d \rbe
\ee
belongs to the same class. Taking the square  we find
\be
\left(\Sigma^C_{\alp\bet}\right)^2 = \oint_C  d \rbe \oint_C  d r'_{\bet}
\ral r'_{\alp} = - \oh  \oint_C  d \rbe \oint_C  d r'_{\bet} (r-r')^2
\ee
where the last transformation follows from the fact, that only the cross
term in $ (r-r')^2$ yields nonzero after double loop integration.

Any expansion in terms  of the square of the tensor area reduces,
therefore to the superposition of multiple loop integral of the product of
$(r_i-r_j)^2 $, which is an example of the abelian Stokes functional.  In
the limit of large area, this could reduce to the fractional power. An
example could be, say
\be
\Psi[C] \stackrel{?}= \EXP{B\left(1-\left(1+ \frac{\et
\left(\Sigma^C_{\alp\bet}\right)^2}{\nu^3} \right)^{\ot}\right)}
\ee
One could explicitly verify all the properties of the abelian Stokes
functional. This example is not realistic, though, as it does not have the
odd terms of expansion. In the real world such terms are present at the
viscous scales. According to our solution, this asymmetry disappears in
inertial range of loops (which does not apply to velocity correlators at
inertial range, as those correspond to shrunk loops).

\section{Matrix Model}

The Navier-Stokes equation represents a very special case of nonlinear
PDE. There is a well known galilean invariance
\begin{equation}
v_{\alpha}(r,t) \rightarrow v_{\alpha}(r-u t,t) + u_{\alpha}
\end{equation}
which relates the magnitude of velocity field with the scales of time
and space. \footnote{At the same time it tells us that the constant part of
velocity if frame dependent, so that it better be eliminated, if we
would like to have a smooth limit at large times. Most of notorious
large scale divergencies in turbulence are due to this unphysical
constant part.} Let us make this relation more explicit.

First, let us introduce the vorticity field
\begin{equation}
\omega_{\mu\nu} = \partial_{\mu} v_{\nu} -\partial_{\nu} v_{\mu}
\end{equation}
and rewrite the Navier-Stokes equation as follows
\begin{equation}
\dot{v}_{\alpha} = \nu \partial_{\beta} \omega_{\beta \alpha} -
v_{\beta}\omega_{\beta \alpha} - \partial_{\alpha} w \\;\;
w = p + \frac{v^2}{2}
\end{equation}

This $ w $ is the well known enthalpy density, to be found from the
incompressibility condition $ div v = 0 $, i.e.
\begin{equation}
\partial^2 w = \partial_{\alpha}v_{\beta}\omega_{\beta \alpha}
\end{equation}

As a next step, let us introduce "covariant derivative" operator
\begin{equation}
D_{\alpha} = \nu \partial_{\alpha} - \frac{1}{2}v_{\alpha}
\end{equation}
and observe that
\begin{equation}
2 \left[D_{\alpha} D_{\beta} \right] = \nu \omega_{\beta \alpha}
\end{equation}
\begin{equation}
2  D_{\beta}\left[ D_{\alpha}D_{\beta} \right] + {\it h.c.}=
\nu \partial_{\beta} \omega_{\beta \alpha} -
v_{\beta}\omega_{\beta \alpha}
\end{equation}
where $ {\it h.c.}$ stands for hermitean conjugate.

These identities allow us to write down the following dynamical
equation for the covariant derivative operator
\begin{equation}
\dot{D}_{\alpha} =  D_{\beta}\left[ D_{\alpha}D_{\beta} \right]
- D_{\alpha} W + {\it h.c.}
\end{equation}

As for the incompressibility condition, it can be written as follows
\begin{equation}
\left[D_{\alpha} D^{\dagger}_{\alpha} \right] =0
\end{equation}
The enthalpy operator $ W = \frac{w}{\nu}$ is to be determined from
this condition , or, equivalently
\begin{equation}
\left[D_{\alpha} \left[D_{\alpha} W \right] \right] =
    \left[D_{\alpha}, D_{\beta}\left[ D_{\alpha}D_{\beta} \right]\right]
\end{equation}

We see, that the viscosity disappeared from these equations. This
paradox is resolved by extra degeneracy of this dynamics: the
antihermitean part of the $ D $ operator is conserved. Its value at
initial time is proportional to viscosity.

The operator equations are invariant with respect to the time
independent unitary transformations
\begin{equation}
D_{\alpha} \rightarrow S^{\dagger}D_{\alpha}S\\;\; S^{\dagger}S = 1
\end{equation}
and, in addition, to the  time dependent unitary
transformations with
\begin{equation}
S(t) = \exp \left( \frac{1}{2\nu} t u_\beta \left(D_{\beta} -
D_{\beta}^{\dagger} \right) \right)
\end{equation}
corresponding to the galilean transformations.

One could view the operator $ D_{\alpha} $ as the  matrix
\begin{equation}
\left\langle i | D_{\alpha} | j \right \rangle =
\int d^3r \psi_i^{\star}(r) \nu \d_{\alp} \psi_j(r)-
\frac{1}{2} \psi_i^{\star}(r)v_{\alpha}(r) \psi_j(r)
\end{equation}
where the functions $ \psi_j(r) $ are the Fourier of Tchebyshev
functions depending upon  the geometry of the problem.

The finite mode approximation would correspond to truncation of this
infinite size matrix to finite size $ N $. This is not quite the same
as leaving $ N $ terms in the mode expansion of velocity field.
The number of independent parameters here is $ O(N^2) $ rather then $
O(N)$. It is not clear whether the unitary symmetry is worth paying
such a high price in numerical simulations!

The matrix model of Navier-Stokes equation has some theoretical beauty
and raises hopes of simple asymptotic probability distribution. The
ensemble of random hermitean matrices was recently applied to the
problem of Quantum Gravity \cite{QG}, which led to a genuine
breakthrough in the field.

Unfortunately, the model of several coupled random matrices, which is
the case here, is much more complicated then the one matrix model
studied in Quantum Gravity. The dynamics of the eigenvalues is coupled
to the dynamics of the "angular" variables, i.e. the unitary matrices
$ S $ in above relations. We could not directly apply the technique of
orthogonal polynomials, which was so successful in the one matrix
problem.

Another technique, which proved to be successful in QCD and Quantum
Gravity is the loop equations. This method, which we are discussing
at length in this paper, works in field theory problems with hidden
geometric meaning. The turbulence proves to be an ideal case, much simpler
then QCD or Quantum Gravity.

\section{The Reduced Dynamics}

Let us now try to  reproduce the dynamics of the loop field by a simpler
Anzatz
\begin{equation}
  \Psi[C] = \left \langle \exp
	\left(
	   \frac{\i}{\nu}\oint d C_{\alpha}(\theta) P_{\alpha}(\theta)
	\right) \right \rangle \label{Reduced}
\end{equation}
The difference with original definition (\ref{eq4}) is that our new
function $ P_{\alpha}(\theta) $ depends directly on $ \theta $ rather
then through the function $ v_{\alpha}(r) $ taken at $ r_{\alpha} =
C_{\alpha}(\theta) $. This is the $ d \rightarrow 1 $ dimensional
reduction we mentioned before. From the point of view of the loop
functional there is no need to deal with field $ v(r) $ , one could
take a shortcut.

Clearly, the reduced dynamics  must be fitted to the Navier-Stokes
dynamics of original field. With the loop calculus, developed above, we
have all the necessary tools to build this reduced dynamics.

Let us assume some unknown dynamics for the $P $ field
\begin{equation}
  \dot{P}_{\alpha}(\theta) = F_{\alpha}\left(\theta,[P] \right)
\end{equation}
and compare the time derivatives of original and reduced Anzatz. We
find in (\ref{Reduced}) instead of (\ref{Orig})
\begin{equation}
     \frac{\i}{\nu}\oint d C_{\alpha}(\theta)
	F_{\alpha}\left(\theta,[P]\right)
\end{equation}

Now we observe, that $P'$ could be replaced by the functional
derivative, acting on the exponential in (\ref{Reduced}) as follows
\begin{equation}
  \frac{\delta}{\delta C_{\alpha}(\theta)}
	\leftrightarrow  -\i\nu P'_{\alpha}(\theta)
\end{equation}
This means, that one could take the operators of the
Section 2, expressing velocity and vorticity in terms of the spike
operator, and replace the functional derivative as above.
This yields the following formula for the spike derivative
\begin{equation}
  D_{\alpha}(\theta,\epsilon) =  -\i\nu \int_{\theta}^{\theta+2
\epsilon} d \phi
	\left(
	1- \frac{\left|\theta + \epsilon - \phi \right|}{\epsilon}
	\right) P'_{\alpha}(\phi) =  -\i\nu
 \int_{-1}^{1}d \mu
	 \mbox{ sgn}(\mu)
	 P_{\alpha} \left(\theta + \epsilon (1+ \mu) \right)
	 \label{DP}
\end{equation}
This is the weighted discontinuity of the function $ P(\theta) $,
which in the naive limit $ \epsilon \rightarrow 0 $ would become the
true discontinuity. However, the function $ P(\theta) $ has in general
the stronger singularities, then discontinuity, so that this limit
cannot be taken yet.

Anyway, we arrive at the dynamical equation for the $P$ field
\begin{equation}
  \dot{P}_{\alpha} = \nu D_{\beta} \Omega_{\beta \alpha} - V_{\beta}
\Omega_{\beta \alpha} \label{Pdot}
\end{equation}
where the operators $ V , D, \Omega $ of the Section 2 should
be regarded as the ordinary numbers, with  definition (\ref{DP}) of $D$ in
terms of $P$.

All the  functional derivatives are gone! We needed them only to
prove equivalence of reduced dynamics to the Navier-Stokes dynamics.

The function $ P_{\alpha}(\theta) $ would become complex now,
as the right side of the reduced dynamical
equation is complex for real $ P_{\alpha}(\theta) $.

Let us discuss this puzzling issue in more detail. The origin of
imaginary units was the factor  of $ \imath $ in exponential of the
definition of the loop field. We had to insert this factor to make the
loop field decreasing at large loops as a result of oscillations of
the phase factors. Later this factor propagated to the definition of
the $ P $ field.

Our spike derivative $ D $ is purely imaginary for real $ P $, and so
is our $ \Omega $ operator. This makes the velocity operator $ V $ real.
Therefore the $ D \Omega $ term in the
reduced equation (\ref{Pdot}) is real for real $ P $ whereas the
$ V \Omega $ term  is purely imaginary.

This does not contradict the moments equations, as we saw before. The
terms with even/odd number of velocity fields in the loop functional are
real/imaginary, but the moments are real, as they should be. The complex
dynamics of $ P $ simply doubles the number of independent variables.

There is one serious problem, though. Inverting the spike operator $
D_{\alpha} $ we implicitly assumed, that it was antihermitean, and
could be regularized by adding infinitesimal negative constant to $
D_{\alpha}^2 $ in denominator. This, indeed, works  perturbatively, in
each term of expansion in time, or that in size of the loop, as we checked.
However, beyond this expansion there would be a problem of
singularities, which arise when $ D_{\alpha}^2(\theta) $ vanishes at
some  $ \theta $.

In general, this would occur for complex $ \theta $, when the
imaginary and real part of $ D_{\alpha}^2(\theta) $ simultaneously
vanish. One could introduce the complex variable
\begin{equation}
  	e^{\imath \theta}=z\\;\;
	e^{-\imath \theta}= \frac{1}{z}\\;\;
	 \oint d \theta = \oint
	\frac{dz}{\imath z}
\end{equation}
where the contour of $z $ integration encircles the origin around the
unit circle.  Later, in course of time evolution, these contours must
be deformed, to avoid  complex roots of $ D_{\alpha}^2(\theta) $.

\section{Initial Data}
Let us study the relation between the initial data for the original
and reduced dynamics. Let us assume, that initial field is distributed
according to some translation invariant probability distribution,
so that initial value of the loop field does not depend on the
constant part of $C(\theta)$.

One can expand translation invariant loop field in functional Fourier
transform
\begin{equation}
  \Psi[C] = \int DQ\delta^3 \left(\oint d \phi Q(\phi) \right)
	 W[Q] \exp
	\left(
	\imath \oint d \theta C_{\alpha}(\theta) Q_{\alpha}(\theta)
	\right)
\end{equation}
which can be inverted as follows
\begin{equation}
  \delta^3 \left( \oint d \phi Q(\phi)\right) W[Q] =
	\int DC\Psi[C]\exp
	\left(
	-\imath \oint d \theta C_{\alpha}(\theta) Q_{\alpha}(\theta)
	\right)
\end{equation}

Let us take a closer look at these formal transformations. The
functional measure for these integrations is defined according to the
scalar product
\begin{equation}
  (A,B) = \oint \frac{d \theta}{2 \pi} A(\theta) B(\theta)
\end{equation}
which diagonalizes in the Fourier representation
\begin{equation}
  A(\theta) = \sum_{-\infty}^{+\infty} A_n e^{\imath n \theta}
\\;\;A_{-n} = A_n^{\star}
\end{equation}
\begin{equation}
  (A,B) =  \sum_{-\infty}^{+\infty} A_n B_{-n} =
A_0 B_0 + \sum_{1}^{\infty} a'_n b'_n + a''_n b''_n\\;\;
a'_n = \sqrt{2} \Re A_n,a''_n = \sqrt{2} \Im A_n
\end{equation}

The corresponding measure is given by an infinite product of the
Euclidean measures for the imaginary and real parts of each Fourier
component
\begin{equation}
  DQ = d^3 Q_0 \prod_{1}^{\infty} d^3 q'_n d^3 q''_n
 \end{equation}
The orthogonality of Fourier transformation could now be explicitly
checked, as
\begin{eqnarray}
  \lefteqn{\int DC \exp \left( \imath \int d \theta C_{\alpha}(\theta)
	\left(
	 A_{\alpha}(\theta) - B_{\alpha}(\theta)
	\right) \right)
	}\\ \nonumber
  &=& \int d^3 C_0 \prod_{1}^{\infty} d^3 c'_n d^3 c''_n \exp
\left( 2 \pi \imath
	\left(
	 C_0 \left(A_0-B_0 \right) +
	\sum_{1}^{\infty} c'_n\left(a'_n - b'_n \right)+
	 c''_n\left(a''_n - b''_n \right)
	\right)
\right)\\ \nonumber
&=&\delta^3\left(A_0-B_0 \right)
\prod_{1}^{\infty} \delta^3\left(a'_n - b'_n \right)
\delta^3\left(a''_n - b''_n \right)
\end{eqnarray}

Let us now check the parametric invariance
\begin{equation}
  \theta \rightarrow f(\theta)\\;\; f(2\pi) -f(0)
= 2\pi \\;\; f'(\theta) >0
\end{equation}
The functions $ C(\theta) $ and $ P(\theta) $ have
zero dimension in a sense, that only their argument transforms
\begin{equation}
  C(\theta) \rightarrow C \left( f(\theta) \right) \\;\;
 P(\theta) \rightarrow P\left( f(\theta) \right)
\end{equation}
The functions $ Q(\theta) $ and $ P'(\theta) $ in above transformation
have dimension one
\begin{equation}
  P'(\theta) \rightarrow  f'(\theta) P'\left( f(\theta) \right)\\;\;
Q(\theta) \rightarrow  f'(\theta) Q \left( f(\theta) \right)
\end{equation}
so that the constraint on $ Q $ remains invariant
\begin{equation}
  \oint d \theta Q(\theta) = \oint df(\theta) Q\left( f(\theta)
\right)
\end{equation}

The invariance of the measure is easy to check for infinitesimal
reparametrization
\begin{equation}
  f(\theta) = \theta + \epsilon(\theta)\\;\; \epsilon(2\pi) = \epsilon(0)
\end{equation}
which changes $C$ and $(C,C)$ as follows
\begin{equation}
  \delta C(\theta) = \epsilon(\theta) C'(\theta) \\;\;
 \delta (C,C) = \oint \frac{d \theta}{2\pi}
\epsilon(\theta) 2 C_{\alpha}(\theta) C'_{\alpha}(\theta) =
-\oint \frac{d \theta}{2\pi}\epsilon'(\theta)C_{\alpha}^2(\theta)
\end{equation}
The corresponding Jacobian reduces to
\begin{equation}
  1 - \oint d \theta \epsilon'(\theta) =1
\end{equation}
in virtue of periodicity.

This proves the parametric invariance of the functional Fourier
transformations. Using these transformations we could find the
probability distribution for the initial data of
\begin{equation}
 P_{\alpha}(\theta) = - \nu\int_{0}^{\theta} d \phi Q_{\alpha}(\phi)
\end{equation}

The simplest but still meaningful distribution of initial velocity
field is the Gaussian one, with energy concentrated in the macroscopic
motions. The corresponding loop field reads
\begin{equation}
  \Psi_0[C] = \exp
	\left(
	 -\frac{1}{2} \oint dC_{\alpha}(\theta)
	\oint dC_{\alpha}(\theta') f\left(C(\theta)-C(\theta')\right)
	\right)
\end{equation}
where $ f(r-r') $ is the velocity correlation function
\begin{equation}
  \left \langle v_{\alpha}(r) v_{\beta}(r') \right \rangle =
\left(\delta_{\alpha \beta}- \partial_{\alpha} \partial_{\beta}
\partial_{\mu}^{-2} \right) f(r-r')
\end{equation}
The potential part drops out in the closed loop integral.

The correlation function varies at macroscopic scale, which means that
we could expand it in Taylor series
\begin{equation}
  f(r-r') \rightarrow f_0 - f_1 (r-r')^2 + \dots \label{Taylor}
\end{equation}
The first term $ f_0 $ is proportional to initial energy density,
\begin{equation}
  \frac{1}{2} \left \langle v_{\alpha}^2 \right \rangle =\frac{d-1}{2}
f_0
\end{equation}
and the second one is proportional to initial energy dissipation
rate
\begin{equation}
 {\cal E}_{0} = -\nu  \left \langle  v_{\alpha} \partial_{\beta}^2
v_{\alpha} \right \rangle = 2 d(d-1) \nu f_1
\end{equation}
where $ d=3 $ is dimension of space.

The constant term in (\ref{Taylor}) as well as $ r^2 + r'^2 $ terms
drop from the closed
loop integral, so we are left with the cross  term $ r r' $
\begin{equation}
  \Psi_0[C] \rightarrow  \exp
	\left(
	 - f_1 \oint dC_{\alpha}(\theta)
	\oint dC_{\alpha}(\theta') C_{\beta}(\theta)C_{\beta}(\theta')
	\right)
\end{equation}
This is almost Gaussian distribution: it reduces to Gaussian one by
extra integration
\begin{equation}
  \Psi_0[C] \rightarrow  {\rm const }\int d^3 \omega \exp
	\left(
	 -\omega_{\alpha \beta}^2
	\right)
	\exp
	\left(
	 2\imath \sqrt{f_1}
	 \omega_{\mu\nu} \oint dC_{\mu}(\theta) C_{\nu}(\theta)
	\right)
\end{equation}
The integration here goes
over all $ \frac{d(d-1)}{2} =3 $ independent $ \alpha < \beta $
components of the antisymmetric tensor $ \omega_{\alpha \beta} $.
Note, that this is ordinary integration, not the
functional one. The physical meaning of this $ \omega $ is the random
constant vorticity at initial moment.

At fixed $ \omega $ the Gaussian functional integration over $ C $
\begin{equation}
  \int DC \exp
	\left(
	 \imath  \oint d \theta
	\left(\frac{1}{\nu}
	  C_{\beta}(\theta) P'_{\beta}(\theta)
	+2 \sqrt{f_1}
	\omega_{\alpha \beta} C'_{\alpha}(\theta)C_{\beta}(\theta)
	\right)
	\right)
\end{equation}
can be performed explicitly, it reduces to solution of the saddle
point equation
\begin{equation}
  P'_{\beta}(\theta) = 4\nu\sqrt{f_1}\omega_{\beta \alpha}
C'_{\alpha}(\theta)
\end{equation}
which is trivial for constant $ \omega $
\begin{equation}
  C_{\alpha}(\theta) =
\frac{1}{4\nu\sqrt{f_1}} \omega^{-1}_{\alpha \beta} P_{\beta}(\theta)
\end{equation}
The inverse matrix is not* unique in odd dimensions, since $ \mbox{Det }
\omega_{\alpha\beta} = 0 $.  However, the resulting pdf for $ P $ is unique.
This is  the Gaussian probability distribution  with the correlator
\begin{equation}
  \left \langle P_{\alpha}(\theta) P_{\beta}(\theta') \right \rangle =
2\imath \nu\sqrt{f_1} \omega_{\alpha \beta} {\rm sign}(\theta'-\theta)
\label{Corr}
\end{equation}

Note, that antisymmetry of $ \omega $ compensates that of the sign
function, so that this correlation function is symmetric, as it should
be. However, it is antihermitean, which corresponds to purely
imaginary eigenvalues. The corresponding realization of the $ P$
functions is complex!

Let us study this phenomenon for the Fourier components.
Differentiating the last equation with respect to $ \theta $ and
Fourier transforming  we find
\begin{equation}
  \left \langle P_{\alpha,n} P_{\beta,m}  \right \rangle
= \frac{4\nu}{m} \delta_{-n m} \sqrt{f_1}\omega_{\alpha \beta}
\end{equation}

This cannot be realized at complex conjugate Fourier components $
P_{\alpha,-n} = P_{\alpha,n}^{\star} $ but we could take
$\bar{P}_{\alpha,n} \equiv P_{\alpha,-n} $ and $ P_{\alpha,n} $ as
real random  variables, with  correlation function
\begin{equation}
  \left \langle \bar{P}_{\alpha,n}P_{\beta,m} \right \rangle
= \frac{4\nu}{m}\delta_{n m}\sqrt{f_1} \omega_{\alpha \beta} \\;\; n>0
\end{equation}
The trivial realization is
\begin{equation}
   \bar{P}_{\alpha,n} =\frac{4\nu}{n} \sqrt{f_1}\omega_{\alpha \beta}
P_{\beta,n}
\end{equation}
with $P_{\beta,n} $ being Gaussian random numbers with unit
dispersion.

As for the constant part $ P_{\alpha,0} $ of $ P_{\alpha}(\theta) $ ,
it is not defined, but it drops from equations in virtue of translational
invariance.

\section{W-functional}

The difficulties of turbulence are hidden in the loop equation, but they
show up, if you try to solve it numerically. The main problem is that
one cannot get rid of the cutoffs $ \epsilon, \delta \rightarrow 0 $
in the definitions of the spike derivatives. These cutoffs are
designed to pick up the singular contributions in the angular
integrals, but with finite number of modes, such as Fourier harmonics
there would be no singularities. We did not find any way to truncate
degrees of freedom in the $ P $ equation, without violating the
parametric invariance. It very well may be, that this invariance would
be restored in the limit of large number of modes, but it looks that
there are too much ambiguity in the finite mode approximation.

After some attempts, we found the simpler version of the loop functional,
which can be studied analytically in the turbulent region. This is
the  generating functional for the scalar products $
P_{\alpha}(\theta_1)P_{\alpha}(\theta_2) $
\begin{equation}
  W[S] = \left \langle \exp
	\left(
	 - \oint d \theta_1 \oint d \theta_2
	S(\theta_1,\theta_2)P_{\alpha}(\theta_1)P_{\alpha}(\theta_2)
	\right) \right \rangle
\label{W1}
\end{equation}
where, as before, the averaging goes over initial data for the $P $
field.

The time derivative of this W-functional
\begin{equation}
  \dot{W} = -2 \left \langle \oint d \theta_1 \oint d \theta_2
	S(\theta_1,\theta_2)P_{\alpha}(\theta_1)\dot{P}_{\alpha}(\theta_2)
	\exp
	\left(
	 - \oint d \theta_1 \oint d \theta_2
	S(\theta_1,\theta_2)P_{\alpha}(\theta_1)P_{\alpha}(\theta_2)
	\right) \right \rangle \label{W2}
\end{equation}
can be expressed in terms of functional derivatives of $W$ by
replacing
\begin{equation}
  P_{\alpha}(\phi_1)P_{\alpha}(\phi_2) \rightarrow
	\frac{\delta}{\delta S(\phi_1,\phi_2)} \label{W3}
\end{equation}
for every scalar product of $P$ fields, which arise after expansion of
the spike derivatives (\ref{DP}), (\ref{OM}), (\ref{VOM}) in the scalar
product
\begin{equation}
  P_{\alpha}(\theta_1)\dot{P}_{\alpha}(\theta_2) =
	 \nu P_{\alpha}(\theta_1) D_{\beta}(\theta_2)
	\Omega_{\beta \alpha}(\theta_2) - P_{\alpha}(\theta_1)
	V_{\beta}(\theta_2)\Omega_{\beta \alpha}(\theta_2)
\label{PP}
\end{equation}

This equation has the structure
\begin{equation}
  \dot{W} = \oint d^2 \theta S(\theta_1,\theta_2)
	\left(
	A_2 \left[\frac{\delta}{\delta S} \right]W +
	A_3 \left[\frac{\delta}{\delta S} \right]D^{-2}(\theta,\epsilon)W
	\right)
	\label{W4}
\end{equation}
where $A_k \left[X \right] $ stands for the $k-$ degree homogenous
functional of
the function $ X(\theta_1,\theta_2) $.

The operator $ D^{-2} $ is also
the homogeneous functional of the negative degree $ k = -1 $. It can
be written as follows
\begin{equation}
  D^{-2}(\theta,\epsilon)W[S] =\int_{0}^{\infty}d \tau W[S + \tau U]
\end{equation}
with
\begin{equation}
  U(\theta_1,\theta_2) = \epsilon^{-2}
 \mbox{ sgn}(\theta+ \epsilon-\theta_1)
 \mbox{ sgn}(\theta+ \epsilon-\theta_2)
\end{equation}

\section{Possible Numerical Implementation}

The above general scheme is fairly abstract and complicated. Could it
lead to any practical computation method? This would depend upon the
success of the discrete approximations of the singular equations of
reduced dynamics.

The most obvious approximation would be the truncation of Fourier
expansion at some large number $ N $. With Fourier components
decreasing only as  powers of $ n $ this approximation is doubtful.
In addition, such truncation violates the parametric invariance which
looks dangerous.

It seems safer to  approximate $ P(\theta) $ by a
sum of step functions, so that it is piecewise constant. The
parametric transformations vary the lengths of  intervals of
constant $ P(\theta) $, but leave invariant these constant values.
The corresponding representation reads
\begin{equation}
  P_{\alpha}(\theta) = \sum_{l=0}^{N} \left(p_{\alpha}(l+1)-p_{\alpha}(l)
\right)
 \Theta \left(\theta-\theta_l \right)\\;\; p(N+1) = p(1),\; p(0) = 0
\label{Thetas}
\end{equation}
It is implied that $ \theta_0 =0 < \theta_1 < \theta_2 \dots < \theta_N <
2\pi $.
By construction, the function $ P(\theta) $ takes value
$p(l)$ at the interval $ \theta_{l-1} < \theta < \theta_{l}
$.

We could take $ \dot{P}(\theta) $ at the middle of this interval as
approximation to $ \dot{p}(l) $.
\begin{equation}
	\dot{p}(l) \approx \dot{P}(\bar{\theta}_l)\\;\;
   \bar{\theta}_l = \frac{1}{2}\left(\theta_{l-1} + \theta_l \right)
\end{equation}
As for the time evolution of angles
$ \theta_l $ , one could differentiate (\ref{Thetas}) in time
and find
\begin{equation}
 \dot{P}_{\alpha}(\theta) =
\sum_{l=0}^{N} \left(\dot{p}_{\alpha}(l+1)-\dot{p}_{\alpha}(l) \right)
 \Theta \left(\theta-\theta_l \right) -
\sum_{l=0}^{N} \left(p_{\alpha}(l+1)-p_{\alpha}(l) \right)
 \delta(\theta-\theta_l)\dot{\theta_l}
\end{equation}
from which one could derive the following approximation
\begin{equation}
\dot{\theta_l} \approx \frac{\left(p_{\alpha}(l)-p_{\alpha}(l+1)
\right)}{ \left(p_{\mu}(l+1)-p_{\mu}(l)\right)^2}
\int_{\bar{\theta}_l}^{\bar{\theta}_{l+1}} d \theta
\dot{P}_{\alpha}(\theta)
\end{equation}

The extra advantage of this approximation is its simplicity. All the
integrals involved in the definition of the spike derivative
(\ref{DP}) are trivial for the stepwise constant $ P(\theta) $. So,
this approximation can be in principle implemented  at the computer.
This formidable task exceeds the  scope of the present  work,
which we view as purely theoretical.

\section{Uniqueness of the tensor area law}

Let us address the issue of the uniqueness of the tensor area solution.
Let us take  the following Anzatz
\be
S[C] = f\left( \oint_C d \ral \oint_C d r'_{\alp} W(r-r') \right)
\ee
When substituted into the static loop equation (with the area derivatives
computed in Appendix A), it yields the following equation for the
correlation function $W(r)$
\bea
0=\oint_C d \ral \oint_C d r'_{\bet} \oint_C d r''_{\gam}
U_{\alp\bet\gam}(r,r',r'') \br
U_{\alp\bet\gam}(r,r',r'')=  W(r-r')\hat{V}^{\alp}_{\bet\gam}W(r'-r'') +
\mbox{permutations}\br
\hat{V}^{\alp}_{\mu\nu} = \del_{\alp\nu} \d_{\mu} -\del_{\alp\mu} \d_{\nu}
\label{U}
\eea
The derivative $f'$ of the unknown function drops from the static equation.

This equation should hold for arbitrary loop $C$. Using the Taylor
expansion for the Stokes type functional \ct{Mig83}, we can argue, that
the coefficient function $U$  must vanish up to the total derivatives. An
equivalent statement is that the third area derivative of this functional
must vanish. Using the loop calculus  (see Appendix A) we find the
following equation
\be
0=\hat{V}^{\alp}_{\mu\nu} \hat{V'}^{\alp'}_{\mu'\nu'}
\hat{V''}^{\alp''}_{\mu''\nu''} U_{\alp\alp'\alp''}(r,r',r'')
\ee
which should hold for arbitrary $r,r',r''$.
This leads to the overcomplete system of equations for $W(r)$ in general
case. However, for the special case $ W(r) = r^2$ which corresponds to the
square of the tensor area
\be
\Sigma_{\alp\bet}^2 = - \oh \oint_C d \ral \oint_C d r'_{\alp}(r-r')^2
\ee
the system is satisfied as a consequence of certain symmetry.  In this
case  we find in the loop equation
\be
2 \oint_C d \ral \oint_C d r'_{\bet}(r-r')^2 \oint_C d r''_{\alp} \left(
r'_{\bet} - r''_{\bet}\right) \propto  \Sigma^C_{\alp\bet}\oint_C d \ral
\oint_C d r'_{\bet}(r-r')^2
\ee
The last integral is symmetric with respect to permutations of $ \alp ,
\bet$, whereas the first  factor $ \Sigma^C_{\alp\bet}$ is antisymmetric,
hence the sum over $\alp\bet$ yields zero, as we already saw above.

It was assumed in above arguments, that the loop $C$  consist of  only one
connected part. Let us now consider the more general situation, with
arbitrary number $n$ of loops $C_1,\dots C_n$. The corresponding Anzatz
would be
\be
S_n\left[C_1,\dots C_n\right] = s_n\left(\Sigma^1,\dots \Sigma^n\right)
\ee
where $\Sigma^i$ are tensor areas.

This function should obey the same WKB loop equations in each variable.
Introducing the loop vorticities
\be
\omega^k_{\mu\nu}  = 2 \pbyp{s_n}{\Sigma^{k}_{\mu\nu}}
\ee
which are constant on each loop, we have to solve the following problem.
What are the values of $ \omega^k_{\mu\nu}$ such that the single velocity
field $\val(r) $ could produce them?

We do not see any other solutions, but the trivial one, with all equal $
\omega^k_{\mu\nu} $ and linear velocity, as before. This would correspond
to
\be
s_n\left(\Sigma^1,\dots \Sigma^n\right)= s_1\left(\Sigma\right)\\;\;
\Sigma_{\mu\nu}=\sum_{k=1}^n \Sigma^k_{\mu\nu}= \oint_{\uplus C_k} r_{\mu}
d r_{\nu}
\ee
The loop equation would be satisfied like before, with $ C = \uplus C_k $.
This corresponds to the additivity of loops
\be
S_n\left[C_1,\dots C_n\right] = S_1\left[\uplus C_k\right]
\ee
Note, that such additivity is the opposite to the statistical
independence, which would imply that
\be
S_n\left[C_1,\dots C_n\right] = \sum S_1\left[C_k\right]
\ee
The  additivity could also be understood as a statement, that any set of
$n$ loops is equivalent to a single loop for the abelian Stokes
functional. Just connect these loops by wires, and note that the
contribution of wires cancels. So, if the area law holds for {\em
arbitrary} single loop, than it must be additive.

This assumption may not be true, though, as it often happens in the WKB
approximation. There is no single asymptotic formula, but rather
collections of different WKB regions, with quantum regions in between. In
our case, this corresponds to the  following situation.

Take the large circular loop, for which the WKB approximation holds, and
try to split it into two large circles. You will have to twist the loop
like the infinity  symbol $ \8$, in which case it intersects itself. At
this point, the WKB approximation might break, as the short distance
velocity correlation might be important near the self-intersection point.
This may explain the paradox of the vanishing tensor area for the $\8$
shaped loop. From the point of view of our area law such loop is not large
at all.

\section{Minimal surfaces}

Let us present here the modern view at the classical theory of the minimal
surfaces. The minimal surface can be described by parametric equation
\be
S: \ral = X_{\alpha}\left(\xi_1,\xi_2\right)
\ee
The function $ X_{\alpha}(\xi) $ should provide the minimum to the
area functional
\be
A[X] = \int_S \sqrt{d \sigma_{\mu\nu}^2} = \int d^2 \xi \sqrt{\mbox{Det } G}
\ee
where
\be
G_{ab} = \d_a X_{\mu}\d_b X_{\mu},
\ee
is the induced metric.
For the general studies it is sometimes convenient to introduce the unit
tangent tensor as an independent field and minimize
\be
A\left[X,t,\lambda\right] = \int d^2 \xi  \left(\,e_{ab} \d_a X_{\mu}
\d_b X_{\nu} \, t_{\mu\nu} + \lambda \left(1-  t_{\mu\nu}^2 \right) \right)
\ee
{}From the classical equations we will find then
\be
t_{\mu\nu} = \frac{e_{ab}}{2\lambda}  \d_a X_{\mu} \d_b X_{\nu}\;;
t_{\mu\nu}^2 = 1,
\ee
which shows equivalence to the old definition.

For the actual computation of the minimal area it is convenient to introduce
the auxiliary internal metric  $ g_{ab} $
\be
A\left[X,g\right] = \oh \int_S d^2 \xi  \, \tr{ g^{-1} G} \,\sqrt{\mbox{Det }
g}.
\label{gG}
\ee
The straightforward minimization with respect to $ g_{ab} $ yields
\be
 g_{ab} \,\tr{ g^{-1} G} =  2 G_{ab},
\ee
which has the family of solutions
\be
g_{ab} = \lambda G_{ab}.
\ee
The local scale factor $ \lambda $ drops from the area functional, and we
recover original definition. So, we could first minimize the quadratic
functional \rf{gG} with respect to $ X(\xi) $ (the linear problem), and
then minimize with respect to $ g_{ab} $ (the nonlinear problem).

The crucial observation is the possibility to choose conformal
coordinates, with the diagonal metric tensor
\be
g_{ab} = \delta_{ab} \rho,\; g^{-1}_{ab} = \frac{\delta_{ab}}{\rho},\;
\sqrt{\mbox{Det } g} = \rho;
\ee
after which the local scale factor $\rho$ drops from the integral
\be
A[X,\rho] = \oh \int_S d^2 \xi \d_a X_{\mu} \d_a X_{\mu}.
\ee
However, the $ \rho $ field is implicitly present in the problem,
through the boundary conditions.

Namely, one has to allow an arbitrary parametrization of the boundary curve
$ C $. We shall use the upper half plane of $ \xi$ for our surface, so the
boundary curve corresponds to the real axis $ \xi_2 = 0 $. The boundary
condition will be
\be
X_{\mu}(\xi_1,+0) = C\left(f(\xi_1)\right),
\label{Bcon}
\ee
where the unknown function $ f(t) $ is related to the boundary value of $
\rho$ by the boundary condition for the metric
\be
g_{11}  = \rho =  G_{11} = \left(\d_1 X_{\mu}\right)^2 = C_{\mu}'^2 f'^2
\ee
As it follows from the initial formulation of the problem, one should now
solve the linear problem for the $ X $ field, compute the area and
minimize it as a functional of $ f(.) $. As we shall see below, the
minimization condition coincides with the diagonality of the metric at the
boundary
\be
\left[\d_1X_{\mu} \d_2 X_{\mu}\right]_{\xi_2=+0} = 0
\label{Diag}
\ee

The linear problem is nothing but the Laplace equation $ \d^2 X = 0 $ in the
upper half
plane with the Dirichlet boundary condition \rf{Bcon}. The solution is
well known
\be
X_{\mu}(\xi) = \int_{-\8}^{+\8} \frac{d t}{\pi} \frac{C_{\mu}\left(f(t)\right)
\,
\xi_2}{\left(\xi_1-t\right)^2 + \xi_2^2}
\ee
The area functional can be reduced to the boundary terms in virtue of the
Laplace equation
\be
A[f]  = \oh \int d^2 \xi \d_a \left(X_{\mu} \d_a X_{\mu}\right) = -\oh
\int_{-\8}^{+\8} d \xi_1 \left[X_{\mu} \d_2  X_{\mu}\right]_{\xi_2 = +0}
\ee
Substituting here the solution for $ X $ we find
\be
A[f] = -\frac{1}{2\pi}\, \Re \int_{-\8}^{+\8}d t \int_{-\8}^{+\8} d t'
\frac{C_{\mu}(f(t))\, C_{\mu}(f(t'))}{(t-t'-\i0)^2}
\ee
This can be rewritten in a nonsingular form
\be
A[f] = \frac{1}{4\pi}\,  \int_{-\8}^{+\8}d t \int_{-\8}^{+\8} d t'
\frac{\left(C_{\mu}(f(t))- C_{\mu}(f(t'))\right)^2}{(t-t')^2}
\ee
which is manifestly positive.

Another nice form can be obtained by integration by parts
\be
A[f] = \frac{1}{2\pi}\,  \int_{-\8}^{+\8}d t f'(t) \int_{-\8}^{+\8} d t'
f'(t') C'_{\mu}(f(t))\, C'_{\mu}(f(t')) \log | t-t'|
\ee
This form allows one to switch to the inverse function $ \tau(f) $ which
is more convenient for optimization
\be
A[\tau] = \frac{1}{2\pi}\,  \int_{-\8}^{+\8}d f \int_{-\8}^{+\8} d f'
 C'_{\mu}(f)\, C'_{\mu}(f') \log |\tau(f)-\tau(f')|
\ee

In the above formulas it was implied that $ C(\8) = 0 $. One could switch
to more traditional circular parametrization by mapping the upper half
plane inside the unit circle
\be
\xi_1 + \i \xi_2 = \i \frac{1-\omega}{1+\omega} \;; \omega = r e^{\i
\alpha}\; ; r \le 1.
\ee
The real axis is mapped at the unit circle. Changing variables in above
integral we find
\be
X_{\mu}(r,\alpha) =  \Re \int_{-\pi}^{\pi}  \frac{d \theta}{\pi}
C_{\mu}(\phi(\theta))
\left(\frac{1}{1- r\EXP{\i\alpha- \i\theta}}- \frac{1}{1 +
\EXP{-\i\theta}}\right)
\ee
Here
\be
\phi(\theta) = f\left(\tan{\frac{\theta}{2}}\right).
\ee
The last term represents an irrelevant translation of the surface, so it
can be dropped. The resulting formula for the area reads
\be
A[\phi] = \frac{1}{4 \pi} \int_{-\pi}^{\pi} d \theta \int_{-\pi}^{\pi} d
\theta'
\frac{\left(C_{\mu}(\phi(\theta))-
C_{\mu}(\phi(\theta'))\right)^2}{\left| e^{\i \theta} - e^{\i \theta'}
\right|^2}
\ee
or, after integration by parts and inverting parametrization
\be
A[\theta] = \frac{1}{2 \pi} \int_{-\pi}^{\pi} d \phi \int_{-\pi}^{\pi} d \phi'
C'_{\mu}(\phi)\,C'_{\mu}(\phi') \log \left| \sin \frac{\theta(\phi) -
\theta(\phi')}{2} \right|
\ee

Let us now minimize the area as a functional of the boundary
parametrization $ f(t) $ (we shall stick to the upper half plane). The
straightforward variation yields
\be
0 = \Re \int_{-\8}^{+\8} d t' \frac{C_{\mu}(f(t'))\,
C_{\mu}'(f(t))}{(t-t'+\i 0)^2}
\label{NL}
\ee
which duplicates the above diagonality condition \rf{Diag}.
Note that in virtue of this condition the normal vector $ n_{\mu}(x) $ is
directed towards $ \d_2 X_{\mu} $ at the boundary. Explicit formula reads
\be
n_{\mu}\left(C(f(t))\right) \propto \Re \int_{-\8}^{+\8} d t'
 \frac{C_{\mu}(f(t'))}{(t-t'+\i 0)^2}
\ee

Let us have a closer look at the  remaining nonlinear integral equation
\rf{NL}. In terms of inverse parametrization it reads
\be
0 = \Re\int_{-\8}^{+\8} d f \frac{ C'_{\mu}(f) C'_{\mu}(f')}{\tau(f)
-\tau(f') + \i 0}
\ee
Introduce the vector set of  analytic   functions
\be
F_{\mu}(z) =
\int_{-\8}^{+\8}\frac{d f}{\pi} \frac{C'_{\mu}(f)}{\tau(f)-z}
\ee
which decrease as $ z^{-2} $ at infinity. The discontinuity  at the real
axis
\be
\Im F_{\mu}(\tau + \i 0) = C_{\mu}'(f) f'(\tau)
\ee
Which provides the implicit equation for the parametrization $ f(\tau) $
\be
\int d \tau \Im F_{\mu}(\tau +\i0) = C_{\mu}(f)
\ee
We see, that the imaginary part points in the tangent direction at the
boundary. As for the boundary value of the real part of $ F_{\mu}(\tau) $ it
points in the normal direction along the surface
\be
\Re F_{\mu} \propto n_{\mu}
\ee
Inside the surface there is no direct relation between the derivatives of $
X_{\mu}(\xi) $ and $ F_{\mu}(\xi) $.

The integral  equation \rf{NL} reduces to the trivial boundary condition
\be
 F_{\mu}^2(t+\i0) =  F_{\mu}^2(t-\i0)
\ee
In other words, there should be no discontinuity of $ F_{\mu}^2 $ at the real
axis.
The solution compatible with analyticity in the upper half plane and $
z^{-2} $ decrease at infinity is
\be
F_{\mu}^2(z) = (1+ \omega)^4\,P(\omega);\; \omega = \frac{\i - z}{\i + z}
\ee
where $ P(\omega) $ defined by a series, convergent at $ |\omega| \le 1 $.
In particular this could be a polynomial.
The coefficients of this series should be found from an algebraic
minimization problem, which cannot be pursued forward in general case.

The flat loops are trivial though.
In this case the problem reduces to the conformal transformation mapping
the loop onto the unit circle. For the unit circle we have simply
\be
C_1 + \i C_2 = \omega;\; F_1 = \i F_2 = - \frac{(1+ \omega)^2}{2};\; P = 0.
\ee
Small perturbations around  the circle or any other flat loop can be
treated in a systematic way, by a perturbation theory.

\section{Kolmogorov triple correlation and time reversal}

Are there any restrictions on the circulation PDF from the known asymmetry of
velocity correlations, in particular, the Kolmogorov triple correlation?
The answer is that the Kolmogorov correlation does not imply the asymmetry
of {\em vorticity} correlations.

Taking the tensor version of the $\frac{4}{5}$ law in arbitrary  dimension
$d$
 \be
\left\langle  v_{\alpha}(0) v_{\beta}(0) v_{\gamma}(r)\right\rangle =
\frac{{\cal E }}{(d-1)(d+2)}
	\left(
	 \delta_{\alpha \gamma} r_{\beta} +
	 \delta_{\beta \gamma} r_{\alpha} -
	 \frac{2}{d}\delta_{\alpha \beta} r_{\gamma}
	\right)
\label{KOLM}
\ee
and differentiating, we find that
\be
\VEV{\val(0)\vbe(0)\omega_{\gam\lam}(r)} = 0
\ee
So, the odd vorticity correlations could, in fact, be absent, in spite of
the asymmetry of the velocity distribution.

\end{document}